\documentclass[twoside,11pt]{article}

\usepackage[accepted, specialissue]{melba} 

\usepackage{amsmath,amsfonts}
\usepackage{graphicx,subcaption}
\usepackage{caption}
\usepackage{appendix}
\usepackage{multirow}
\usepackage{xcolor}
\usepackage{svg}

\melbaid{2023:015}  
\doi{https://doi.org/10.59275/j.melba.2023-b7bc}
\melbaauthors{Ramirez et al.}  
\volume{2}
\firstpageno{406}  
\melbayear{2023}  
\datesubmitted{04/2023}  
\datepublished{10/2023}  

\melbaspecialissue{Perinatal, Preterm and Paediatric Image Analysis (PIPPI) 2022}

\melbaspecialissueeditors{Jana Hutter, Roxane Licandro, Andrew Melbourne, Esra Abaci Turk, \\ Daphna Link-Sourani, Christopher Macgowan}

\ShortHeadings{Multi-task learning for anomaly segmentation}{Ramirez et al.}

\title{Multi-task learning for joint weakly-supervised segmentation and aortic arch anomaly classification in fetal cardiac MRI}

\author{\name Paula Ramirez \orcid{0000-0002-0705-5296} \email paula.ramirez\_gilliland@kcl.ac.uk  \AND 
\name Alena Uus \orcid{0000-0001-5796-2145} \email alena.uus@kcl.ac.uk  \AND
\name Milou P.M. van Poppel \orcid{0000-0002-1739-4726} \email milou.van\_poppel@kcl.ac.uk  \AND
\name Irina Grigorescu \orcid{0000-0002-9756-3787} \email irina.grigorescu@kcl.ac.uk  \AND
\name Johannes K. Steinweg \orcid{0000-0002-3366-0932} \email johannes.steinweg@kcl.ac.uk  \AND
\name David F.A. Lloyd \orcid{0000-0003-1759-6106} \email david.lloyd@kcl.ac.uk  \AND
\name Kuberan Pushparajah \orcid{0000-0003-1541-1155} \email kuberan.pushparajah@kcl.ac.uk \AND
\name Andrew P. King \orcid{0000-0002-9965-7015} \email andrew.king@kcl.ac.uk \AND
\name Maria Deprez \orcid{0000-0002-2799-6077} \email maria.deprez@kcl.ac.uk \AND
 \addr School of Biomedical Engineering and Imaging Sciences, King's College London, London, UK.
}

\begin{document}

\maketitle

\begin{abstract}
Congenital Heart Disease (CHD) is a group of cardiac malformations present already during fetal life, representing the prevailing category of birth defects globally. Our aim in this study is to aid 3D fetal vessel topology visualisation in aortic arch anomalies, a group which encompasses a range of conditions with significant anatomical heterogeneity. We present a multi-task framework for automated multi-class fetal vessel segmentation from 3D black blood T2w MRI and anomaly classification. Our training data consists of binary manual segmentation masks of the cardiac vessels' region in individual subjects and fully-labelled anomaly-specific population atlases. Our framework combines deep learning label propagation using VoxelMorph with 3D Attention U-Net segmentation and DenseNet121 anomaly classification. We target 11 cardiac vessels and three distinct aortic arch anomalies, including double aortic arch, right aortic arch, and suspected coarctation of the aorta. 
We incorporate an anomaly classifier into our segmentation pipeline, delivering a multi-task framework with the primary motivation of correcting topological inaccuracies of the segmentation. The hypothesis is that the multi-task approach will encourage the segmenter network to learn anomaly-specific features. As a secondary motivation, an automated diagnosis tool may have the potential to enhance diagnostic confidence in a decision support setting. Our results showcase that our proposed training strategy significantly outperforms label propagation and a network trained exclusively on propagated labels. Our classifier outperforms a classifier trained exclusively on T2w volume images, with an average balanced accuracy of 0.99 (0.01) after joint training. Adding a classifier improves the anatomical and topological accuracy of all correctly classified double aortic arch subjects. 

Our code is available at \\ ~\url{https://github.com/SVRTK/MASC-multi-task-segmentation-and-classification}.
\end{abstract}

\begin{keywords}
	Deep Learning, Fetal Cardiac Imaging, Congenital Heart Disease, Automated Diagnosis, Fetal Cardiac MRI, Aortic Arch Segmentation, Multi-Task Learning, Multi-Class Vessel Segmentation, Anomaly Segmentation 
\end{keywords}

\section{Introduction}
\label{sec:introduction}

Congenital heart disease (CHD) is the leading cause of mortality related to congenital defects (\cite{mendis2011global}). Accurate CHD diagnosis before birth is essential to inform appropriate early postnatal management, which is known to lead to improved patient outcomes both in terms of mortality and long-term morbidity (\cite{brown2006delayed, mazwi2013unplanned}). 

We present a fully automated weakly-supervised multi-task tool for multi-class fetal cardiac vessel segmentation and anomaly classification in 3D T2w MRI. Our intention is to facilitate fetal cardiac vessel visualisation for prenatal diagnostic reporting purposes, and provide the groundwork for automated vessel biometry and detection. We target three aortic arch anomalies: Right Aortic Arch (RAA), Double Aortic Arch (DAA), and suspected Coarctation of the Aorta (CoA). 

In current clinical practice, vessel segmentation in fetal cardiac MRI is a time-consuming process based on thresholding followed by manual correction, resulting in a binary mask of fetal heart and vessels. An expert fetal cardiac clinician generally takes 1-2 hrs to complete the binary mask, which is a significant hurdle to wider translation outside the research setting. In addition, more refined multi-label segmentation would provide clinicians with better visualisation of individual vessels, and facilitate automated quantitative analysis, thus reducing reporting time and potentially improving the prediction of outcomes. However, manual multi-label segmentation would bring extra clinical burden and is therefore not currently performed in clinical practice. This is partly due to the image quality available, paired with the small fetal vessel size (often only 1-2 voxels wide).

Our proposed weakly supervised deep learning framework is able to leverage existing binary manual segmentations in conjunction with a small number of condition-specific multi-label atlases, to provide a fully automated and accurate multi-label segmentation of individual cardiac vessels. The technique is adaptable to clinical workflows as it does not require any manual input at inference time. We incorporate an aortic arch anomaly classifier into our framework, with the intention of both providing a clinically useful diagnostic tool and improving segmentation performance.

\subsection{Imaging the fetal heart}

2D Fetal echocardiography is commonly used for detecting CHD before birth, as it offers clear discernment of the cardiac chambers and cardiac functional measurements using Doppler flow. Extracting vessel positional and topological information from 2D echocardiography images alone is an extremely challenging task, requiring highly trained experts. Although 3D STIC \citep{devore2003spatio} is available in many clinical settings, the corruption by fetal motion renders this imaging modality less viable for clinical assessments.

Recently, fetal cardiac MRI has shown the potential as an adjunct to echocardiography in the detection of CHD prenatally \citep{lloyd2019three}, and its use is becoming widespread \citep{dong2018utility,salehi2021utility,dong2020fetal}. State-of-the-art motion correction algorithms \citep{uus2020deformable} address fetal CMR motion challenges, allowing the generation of high-quality 3D MRI of fetal cardiac vessels. T2w black blood 3D MRI offers excellent vascular visualisation \citep{lloyd2019three}, as the vessels appear dark in contrast with surrounding tissue such as the lungs. This is the principal motivation behind the choice of imaging modality used in the present work, given the aim is segmentation of vascular structures in aortic arch anomalies.

\subsection{Deep learning segmentation}

In order to provide a robust assessment of the cardiovascular anatomy in CMR, anatomical structures require segmentation, i.e. voxelwise classification into an anatomical class. Deep learning has become widespread for automated segmentation \citep{hesamian2019deep} and adult CMR segmentation \citep{chen2020deep}.

 U-Net-based architectures \citep{ronneberger2015u,isensee2018nnu} have been successfully employed for fetal brain and thorax MRI segmentation \citep{salehi2018real,uus20213d}. The fetal brain is the most commonly targeted fetal organ for automated segmentation in recent works \citep{keraudren2014automated,khalili2019automatic,ebner2020automated,payette2020efficient,salehi2018real}, due to the growing interest in early brain development. In addition, as a rigid, heterogeneous organ constrained with the fetal skull the brain represents an attractive target for these methods, compared to the fetal heart which regularly deforms with the rapid fetal cardiac cycle. Most forms of CHD also present significant anatomical variation in terms of vessel sizes and relative positions, making the application of the methods particularly challenging.

Deep learning has been extensively applied to adult CMR segmentation in CHD \citep{arafati2019artificial}. \cite{yu20163d} present a 3D fractal network for whole heart and great vessel segmentation, while \cite{rezaei2018whole} propose a framework comprising a three-stage cascade of conditional GANs. \cite{xu2019whole} address the anatomical variability challenge in CHD by employing deep neural networks to segment the myocardial blood pool and chambers, coupled with graph matching for vessel identification. Fetal cardiac datasets, however, differ significantly from those acquired in postnatal life. For example, motion corruption is particularly common in fetal imaging, due to gross fetal, fetal cardiac and maternal respiratory motion. The fetal cardiac structures are also diminutive (most vessels are only millimetres in diameter), and are subject to low signal-to-noise ratio (SNR) and resolution with no option for intravenous contrast-enhancing agents. There is also differing baseline contrast due to the heart being surrounded by fluid filled lungs, as well as issues around temporal resolution due to the rapid fetal cardiac cycle. The technical approaches required for fetal cardiac imaging, such as fast acquisition sequences \citep{patel1997half,semelka1996haste} and high quality slice-to-volume registration algorithms \citep{kuklisova2012reconstruction,uus2020deformable} are therefore highly specialised for this application, and generate unique output data which are not comparable to postnatal imaging methods. A bespoke approach for the application of deep learning methods is therefore required.

\subsection{Atlas-guided segmentation}

\begin{figure}[h]
     \centering
     \begin{subfigure}[b]{0.45\textwidth}
         \centering
         \includegraphics[width=\textwidth]{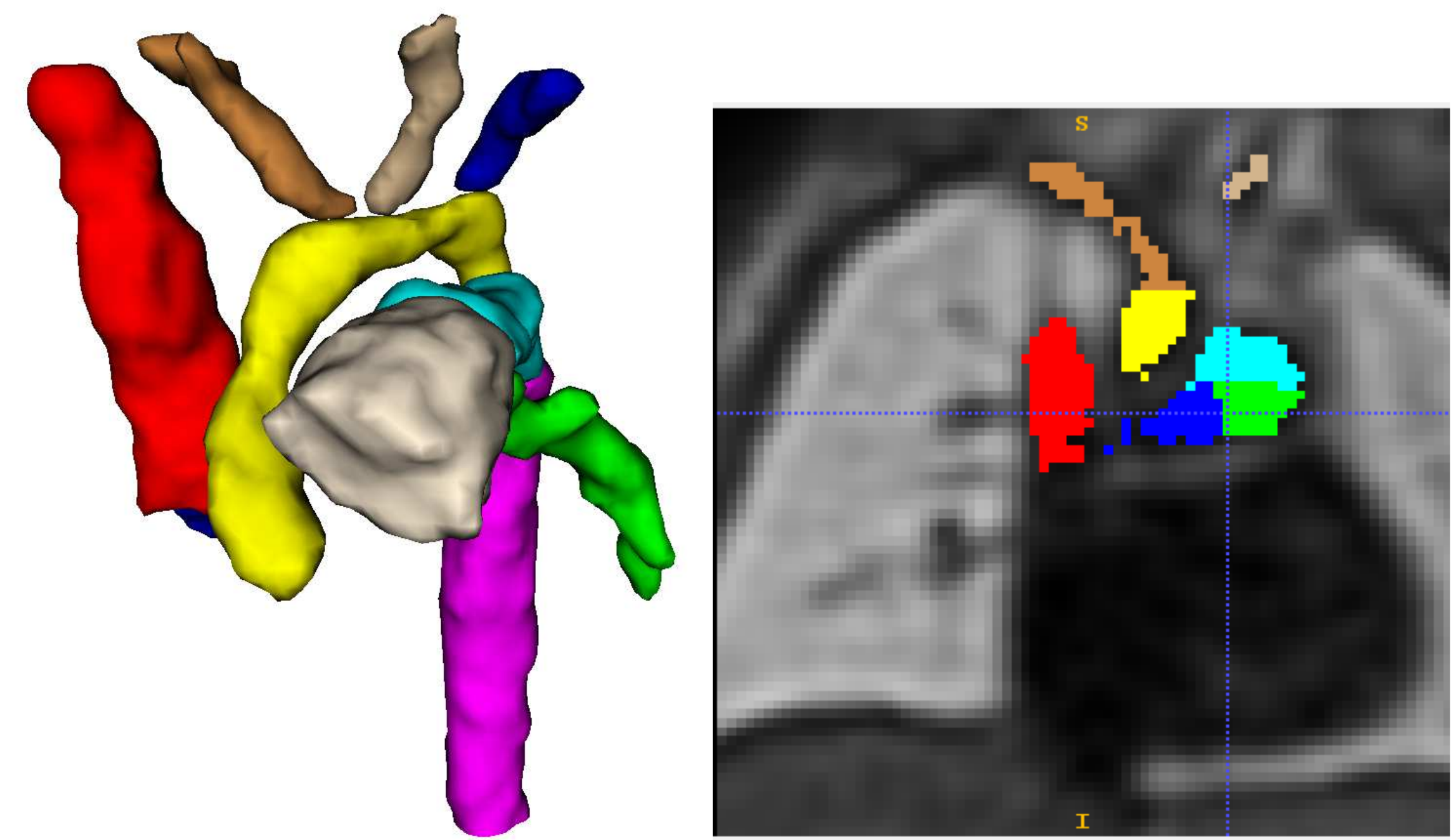}
         \caption{Fully-labelled atlas}
     \end{subfigure}
     \hfill
     \begin{subfigure}[b]{0.45\textwidth}
         \centering
         \includegraphics[width=\textwidth]{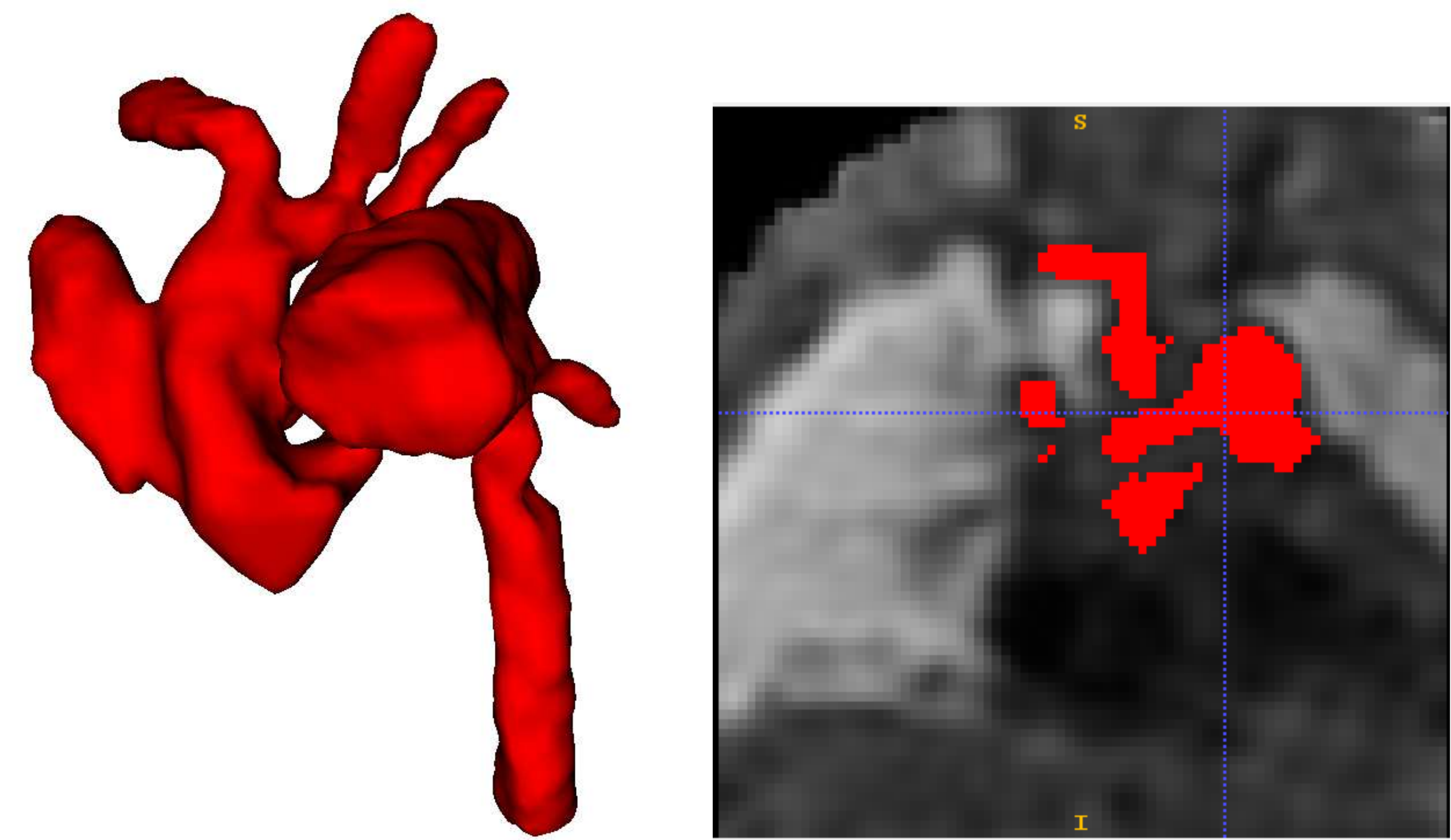}
         \caption{Partially-labelled training subject}
     \end{subfigure}
        \caption{Illustration of our dataset setup.}
        \label{example-data}
\end{figure}

Our training dataset 
consists of partially labelled subjects (binary manually segmented labels) and fully-labelled atlases (Fig.~\ref{example-data}). This type of setup is not uncommon in the medical imaging field, with exhaustive works addressing this challenge \citep{peng2021medical}.

Atlas-based label propagation has been widely used for medical image segmentation \citep{aljabar2009multi,heckemann2006automatic}. 
It uses image registration to transfer labelling information from a given atlas to individual subjects. We propose to use \textbf{VoxelMorph} \citep{balakrishnan2019voxelmorph} for 
propagating multi-class labels from anomaly-specific atlases to fetal subjects.

Deep learning atlas-based segmentation approaches include \cite{dinsdale2019spatial}, which presents a binary mask warping framework based on spatial transformer networks; \cite{xu2019deepatlas} propose to jointly train registration and segmentation networks to tackle partially labelled data. Similarly, in \cite{sinclair2022atlas}, registration and segmentation are performed within the same framework, with the notable addition of a population-derived atlas being constructed in the process. Alternative strategies include one-shot or few-shot segmentation techniques \citep{zhao2019data}, where synthetic labelled data is generated using VoxelMorph and used to deal with partially labelled datasets.

However, label propagation strategies are always limited by registration accuracy, hence in this work we propose to combine the benefits of atlas-based and deep learning based segmentation and use the atlas-propagated labels to train a convolutional neural network (CNN) for the task of fetal heart segmentation.

\subsection{Anomaly classifier}

We also propose to leverage the notable topological distinction between aortic arch anomalies (Fig.~\ref{fig:condition_desc}) to explore the addition of an anomaly classifier into our framework, based on the pipeline presented by \cite{puyol2021fairness} which included a race classifier in combination with a segmentation network. A similar approach has been taken by \cite{mehta2018net}, who address joint segmentation and classification of breast biopsy images, with a U-Net-based network predicting both a standard segmentation and discriminative map which are combined for tissue classification.

The benefits of joint classifier and segmenter network training have also been evidenced in the field of computer vision.~\cite{liao2016understand} propose an approach which in essence is the reverse of our strategy, namely, features from a classifier network are used to train a scene segmentation network. MultiNet (\cite{teichmann2018multinet}) also follows an analogous architecture to ours, combining a shared encoder with a task-specific decoder.

Our primary motivation is to explore whether the addition of a classifier can improve segmentation performance, in particular ensuring the correct topology of the automated segmentation outputs. As a secondary motivation, we explore the possibility automated diagnosis of aortic arch anomalies as a potential aid to clinical decision making.

\subsection{Contributions}
\label{sec:contributions}

In this work, we present a \textbf{weakly supervised multi-task framework for automated multi-class fetal cardiac vessel segmentation and anomaly classification} from 3D reconstructions of T2w black blood MRI images. We propose a deep learning framework, addressing three aortic arch anomalies.

We expand on our prior research \citep{ramirez2022automated}
 by incorporation of an anomaly classifier which improves segmentation discernment between anomalies. We also include detailed ablation studies to validate individual elements of our multi-task framework, which now consists of training using labels propagated from the anomaly-specific atlases and manual labels in individual images, while simultaneously classifying the anomaly from the predicted segmentation. This way 
 we can leverage valuable anomaly-specific features learnt from a classifier to improve our segmentation output. A key novelty is the application of our framework to fetal Cardiac Magnetic Resonance (CMR) data, which is to date largely unexplored. Our framework expands on our previous work \citep{uus20223d}, which for the first time proposed an automated segmentation of fetal cardiac vessels by training a CNN on labels propagated from anomaly-specific atlases.

Our clinical contribution is to aid 3D vessel topology visualisation for clinical reporting and training purposes and to increase confidence in diagnostic accuracy. We propose a multi-class approach to enable isolated visualisation of the anomaly area and affected vessels. Our approach does not require a fully-labelled training set, only subject-specific binary labels alongside multi-class anomaly-specific atlases; and thus is adaptable to clinical environments. These labels are only required for training, segmentation during inference is carried out on fully unlabelled data, from a single network.

\section{Methods}

\subsection{Dataset description}

\subsubsection{Aortic arch anomalies}

\begin{figure}[h]
    \centering
    \includegraphics[width=0.9\textwidth]{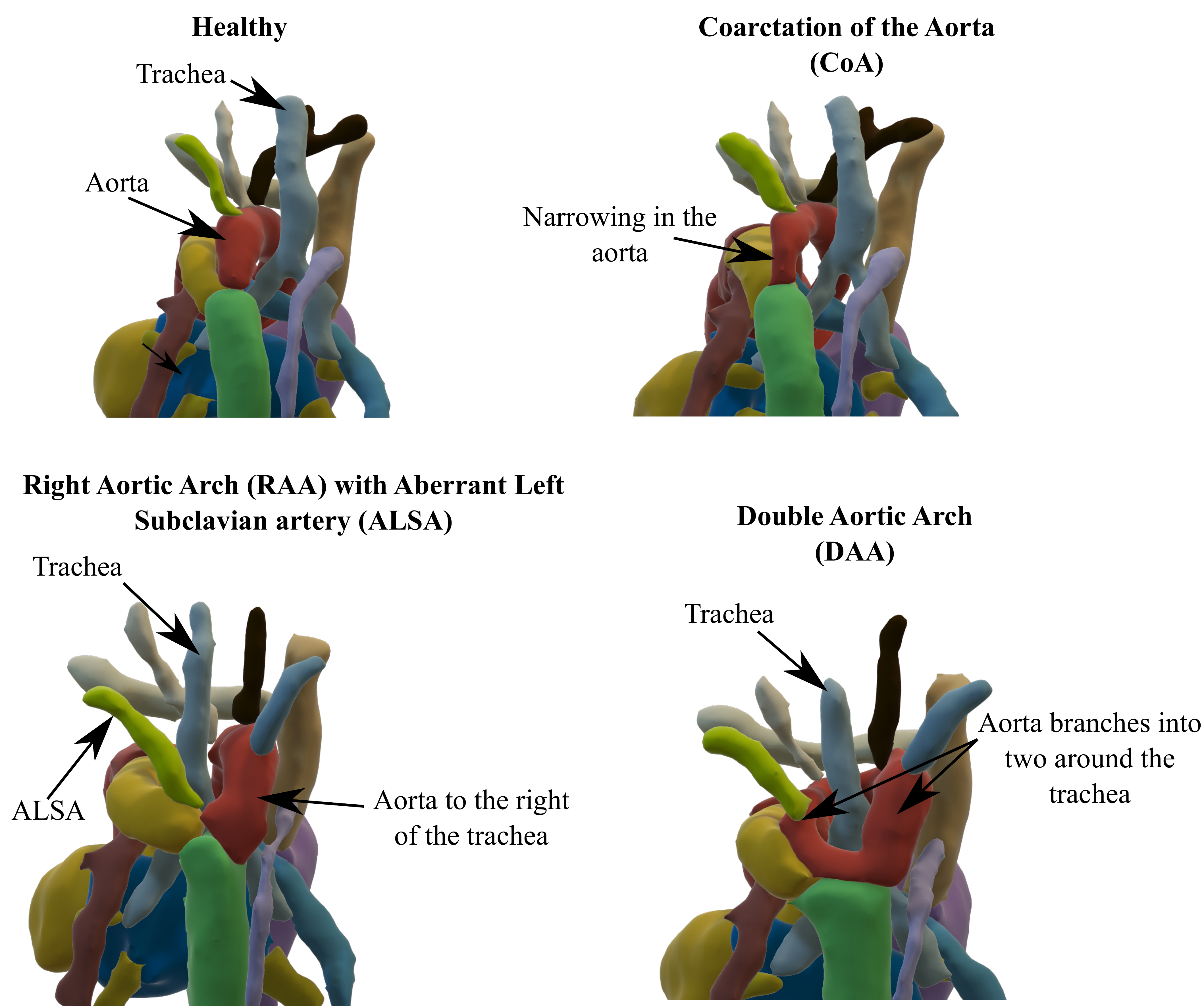}
    \caption{Aortic arch anomaly illustration (atlases by \cite{uus20223d}).}
    \label{fig:condition_desc}
\end{figure}

Our automated segmentation tool is aimed at subjects with Right Aortic Arch (RAA) with aberrant left subclavian artery (ALSA), Double Aortic Arch (DAA), and suspected Coarctation of the Aorta (CoA). These three distinct anomalies are depicted in Fig.~\ref{fig:condition_desc}.

Briefly, in CoA, there is a narrowing of the distal aortic arch which can require urgent postnatal intervention if severe. In RAA the aortic arch passes to the right of the trachea (as opposed to the normal position on the left) with an associated risk of airway and/or oesophageal compression, particularly when associated with an ALSA. In DAA, both right and left aortic arches are present, forming a "vascular ring" around the trachea and oesophagus, which in turn is associated with a higher risk of postnatal symptoms requiring surgical intervention. Whilst each of these categories presents a distinct topological phenotype, there is significant heterogeneity within each group in terms of vessel size and morphology.

Our dataset is representative of patients that are referred to a specialist fetal cardiac service, and therefore no controls are included in this study. Nonetheless, cases with Suspicion of CoA have the same vessel topology as healthy controls.

\subsubsection{Data specifications}
\label{sec:data_specifications}

\begin{figure}[h]
     \centering
     \begin{subfigure}[b]{0.4\textwidth}
         \centering
         \includegraphics[height=5.6cm]{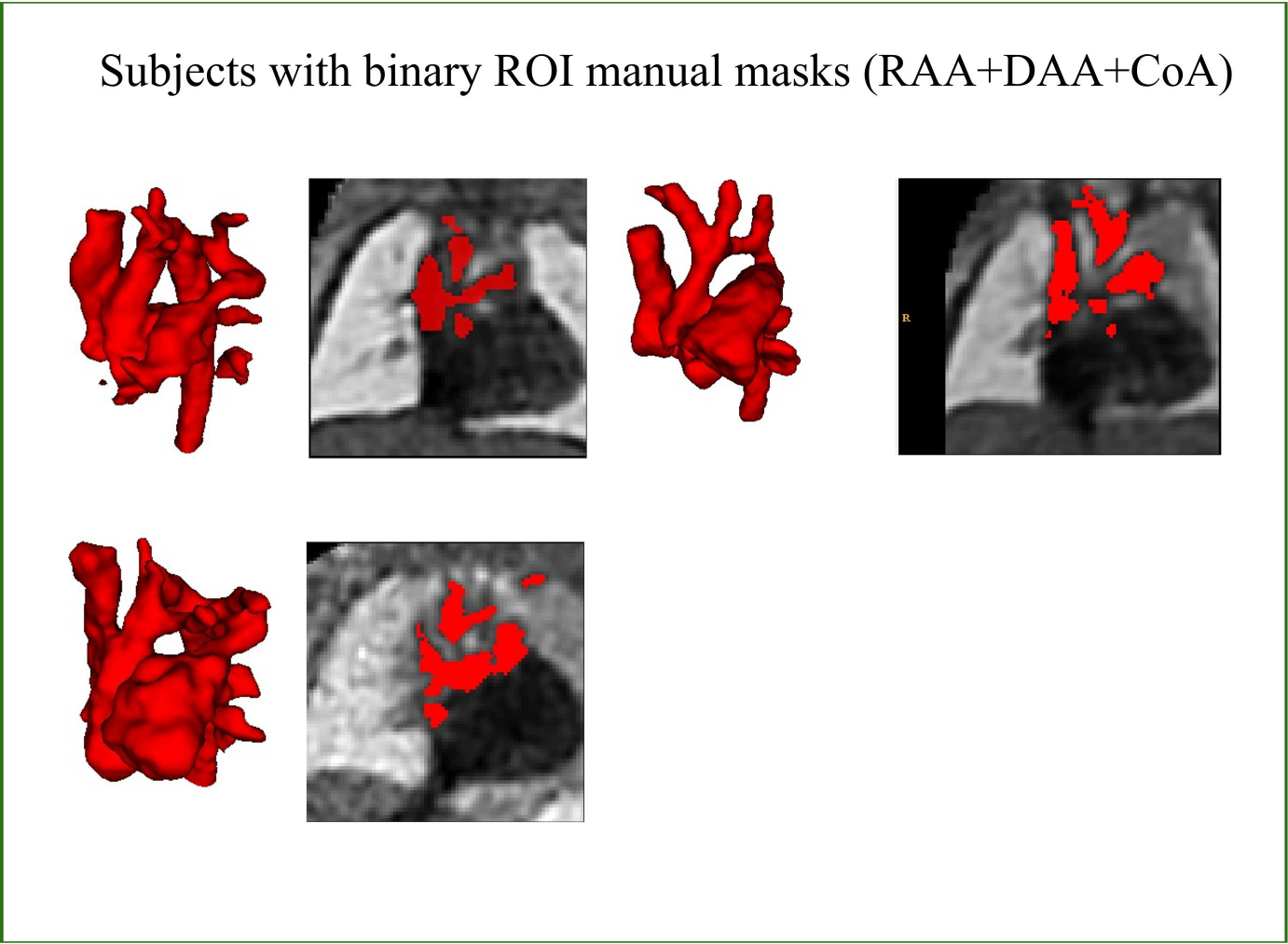}
         \caption{3D reconstructed T2w MRI images with binary manually segmented labels.}
         \label{fig:binary_data}
     \end{subfigure}
     \hfill
     \begin{subfigure}[b]{0.5\textwidth}
         \centering
         \includegraphics[height=5.6cm]{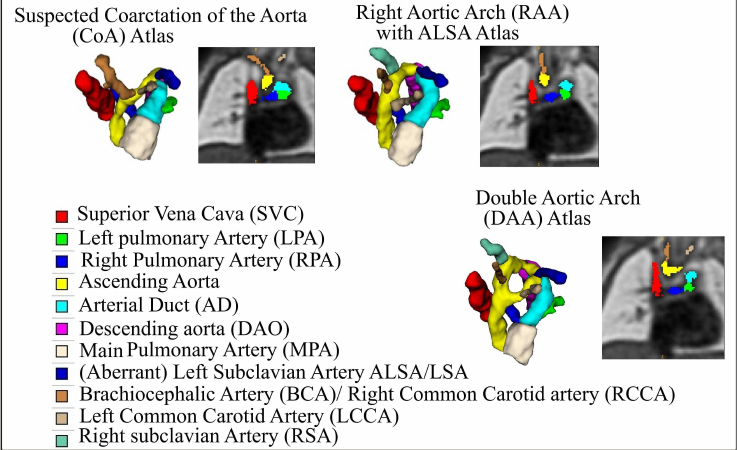}
         \caption{3D T2w MRI aortic arch anomaly atlases with multi-class vessel segmentations}
         \label{fig:atlas_data}
     \end{subfigure}
        \caption{Depiction of our two dataset groups: individual fetal subjects with binary masks (\ref{fig:binary_data}), and multi-class anomaly-specific atlases (\ref{fig:atlas_data})}
        \label{fig:dataset}
\end{figure}

We employ 3D reconstructions of T2w black blood MRI. Our dataset consists of 195 fetal subjects with suspected coarctation of the aorta (CoA, N=94), Right Aortic Arch (RAA) with ALSA (N=72) and Double Aortic Arch (DAA, N=29), 31.4$\pm$1.5 weeks mean gestational age (min=29 weeks, max=36 weeks). In addition to the primary diagnosis of suspected CoA, RAA, or DAA, 56 cases presented secondary diagnoses affecting additional anatomical areas, outside of the segmentation area (aside from RAA$>$DAA). These are included in  Table~\ref{tab:additional_chd}.

\begin{table}[h]
\centering
\caption{Number of cases with an additional diagnosis. AVSD: atrioventricular septal defect, VSD: ventricular septal defect, MV: mitral valve, PAPVD: partial anomalous pulmonary venous drainage, LV/RV: left/right ventricle, AS: aortic valve stenosis, DV: ductus venosus, BAV: bicuspid aortic valve, R/LAD: right/left arterial duct, R/LAA: right/left aortic arch, ?=suspected.}

\begin{tabular}{c|lc}
\textbf{Primary  diagnosis}   & \textbf{Additional   diagnosis}                 & \textbf{Number of cases} \\ \hline
\multirow{9}{*}{Suspected CoA} & (unbalanced) (A)VSD                              & 17                       \\
                               & ?Parachute MV                                  & 1                        \\
                               & Small   pericardial effusion                     & 1                        \\
                               & PAPVD                                            & 1                        \\
                               & LV $<$ RV                                          & 1                        \\
                               & (aneurysmal)   AS                                & 3                        \\
                               & Aberrant DV                                      & 1                        \\
                               & Aneurysm                                         & 2                        \\
                               & (?)BAV                                           & 3                        \\ \hline
\multirow{2}{*}{RAA with ALSA} & \multicolumn{1}{c}{(?) VSD}                      & 4                        \\
                               & \multicolumn{1}{c}{RAD}                          & 1                        \\ \hline
\multirow{4}{*}{DAA}           & \multicolumn{1}{c}{RAA$>$LAA}                      & 10                       \\
                               & \multicolumn{1}{c}{LAD}                          & 9                        \\
                               & \multicolumn{1}{c}{Retro-aortic innominate vein} & 1                        \\
                               & \multicolumn{1}{c}{VSD}                          & 1                       
\end{tabular}
\label{tab:additional_chd}
\end{table}

The datasets were acquired at Evelina London Children's Hospital 
using a 1.5 Tesla Ingenia MRI system and a T2-weighted SSFSE sequence (RT=20,000 ms, ET=50ms, FA=90$^{\circ}$, voxel size=$1.25 \times 1.25$ mm, slice thickness=2.5 mm and slice overlap=1.25 mm). All research participants provided written informed consent. The raw datasets consisted of six to 12 multi-slice stacks of 2D images, covering the fetal thorax in three orthogonal planes.

We use images reconstructed both with Slice-to-Volume Registration (SVR) \citep{kuklisova2012reconstruction, kainz2015fast} and Deformable SVR \citep{uus2020deformable,uus2022media} (DSVR, higher quality, N=49) to 0.75 mm isotropic resolution, to ensure a varied dataset.

Trained clinicians manually segmented a binary vessels label for the majority of our subjects (N=181), Fig.~\ref{fig:binary_data}. We employ propagated atlas labels exclusively for the remaining unsegmented cases (N=14, Sec.~\ref{sec:dl_framework}). In order to achieve our multi-class output, we employ three fully-labelled atlases\footnote{https://gin.g-node.org/SVRTK/} (\cite{uus20223d}, see Fig.~\ref{fig:atlas_data}), one per anomaly (RAA, DAA and CoA). These include 11 manually segmented vascular regions for RAA and DAA, and 10 for CoA cases.

We affinely register all subject images to the pertinent anomaly-specific atlas prior to training via MIRTK\footnote{https://github.com/BioMedIA/MIRTK} \citep{schnabel2001generic,rueckert1999nonrigid,rueckert1999comparison}, and crop to a standardised cardiac vessels region after affine alignment  (see Fig.~\ref{fig:dataset}). We split the subjects into a training set (N$_{CoA}$=71, N$_{RAA}$=54, N$_{DAA}$=21), validation set (N$_{CoA}$=3, N$_{RAA}$=3, N$_{DAA}$=3), and testing set (N$_{CoA}$=20, N$_{RAA}$=15, N$_{DAA}$=5). We normalise and rescale the intensities to between 0 and 1, and use a weighted random sampler to correct for the class imbalance problem.

\subsection{Multi-task segmentation framework}
\label{sec:dl_framework}

Our benchmark framework is based on \cite{ramirez2022automated}, where Attention U-Net \citep{oktay2018attention} is trained using both manual binary labels and multi-class labels propagated from an atlas \citep{uus20223d}. 

\begin{figure*}[h]
    \centering
    \includegraphics[width=\textwidth]{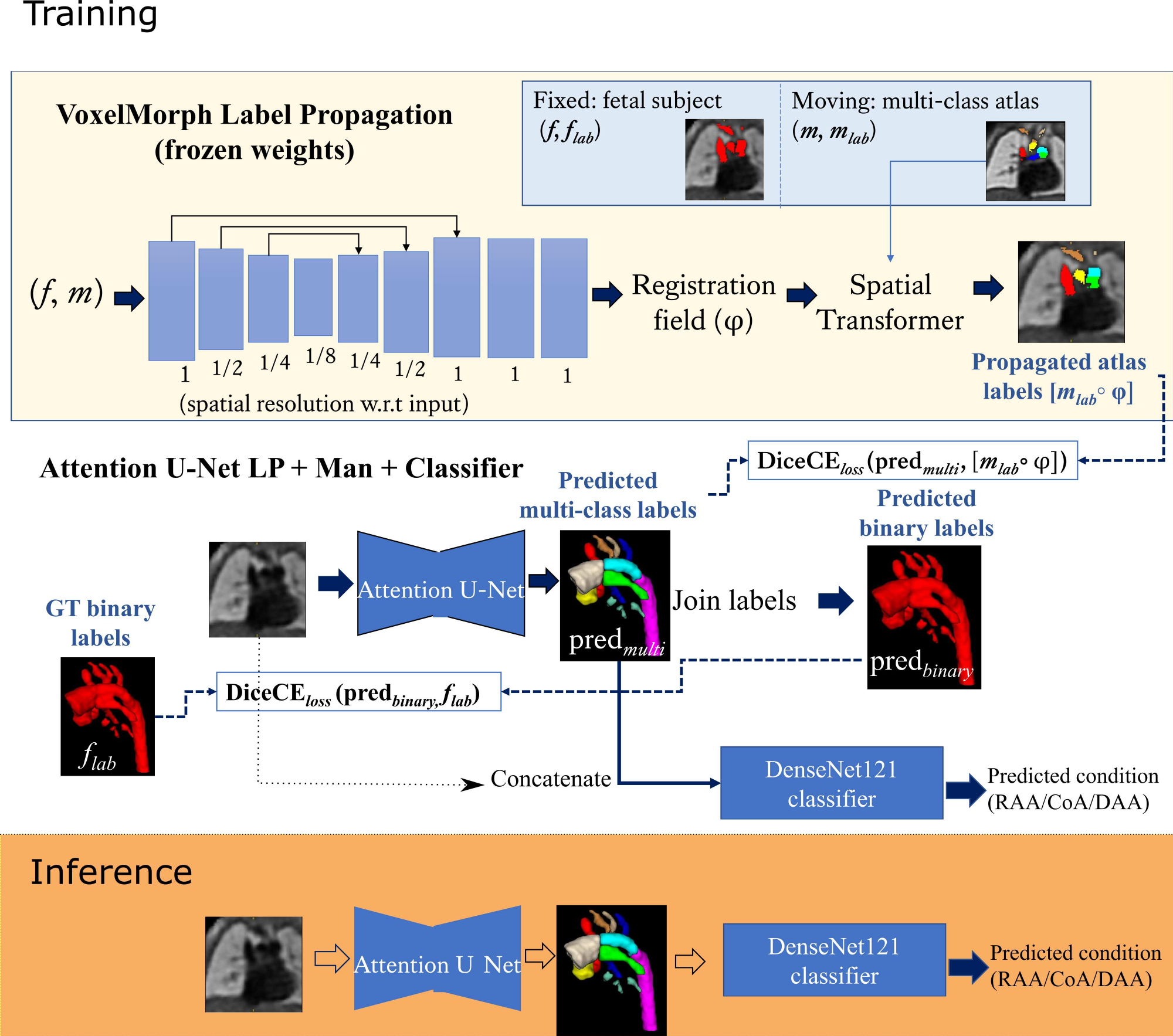}
    \caption{Our proposed segmentation framework (\textit{Attention U-Net LP + Man + Classifier}) is trained with both a multi-class propagated labels loss, and a binary loss employing the manual binary masks. We use three atlases: CoA, RAA and DAA, and train our segmentation network jointly with an anomaly classifier. No labelling information (segmentation or diagnosis) is required during inference. } 

    \label{fig:overall_architecture}
\end{figure*}

The input to the network is an MRI image, and the output is a multi-class segmentation (see Fig.~\ref{fig:overall_architecture}). 
We use VoxelMorph \citep{balakrishnan2019voxelmorph} for label propagation, defining our atlases as moving images ($m$), and fetal subject images as fixed images ($f$).  

Subsequent to VoxelMorph training, we generate our propagated labels and use these to train a CNN for segmentation. We keep the weights of VoxelMorph label propagation frozen while training our segmentation network. We train our proposed network with two losses: (1) a multi-class loss between the propagated labels and the network output; (2) a binary loss between the multi-class network output joined into a binary segmentation and the manual binary labels.

Finally, we attach a DenseNet121 \citep{huang2017densely} classifier to our softmax segmentation output, which predicts one out of three possible diagnoses (CoA, RAA or DAA). We train Attention U-Net and DenseNet121 jointly. This classifier has the option of incorporating the image as an extra channel, to make the classification less dependent on accurate segmentation.

\subsection{Label propagation}
\label{sec:label propagation}

We use VoxelMorph \citep{balakrishnan2019voxelmorph} for label propagation. We warp the labels from the pertinent anomaly-specific atlas (moving images, $m$) into each subject space (fixed images $f$), using the prior anomaly diagnosis knowledge to select the relevant atlas. Note that at testing or inference time, no prior diagnosis knowledge is required, and our framework offers an automated diagnosis prediction.

\subsubsection{Label propagation loss functions}
\label{sec:LP_loss}

Our similarity reconstruction loss function ($\mathcal{L}_{sim}$) is Local Normalised Cross Correlation loss ($\textrm{LNCC}_{loss}$) \citep{balakrishnan2019voxelmorph}. We employ bending energy \textbf{BE loss} ($\mathcal{L}_{BE}$), as described in \cite{grigorescu2020diffusion} as a displacement field regulariser. The \textbf{total registration loss} $\mathcal{L}_{reg}$ may be expressed as 
\begin{equation}
\label{eqn:reg}
    \mathcal{L}_{reg} = \mathcal{L}_{sim}(f,m\cdot \phi) + \lambda_1 \mathcal{L}_{BE} (\phi) ,
\end{equation}
where $\lambda_1$ is a loss weight.

\subsubsection{Registration network implementation details}
\label{sec:voxelmorph_implementation}

\begin{figure}[h]
    \centering
    \includegraphics[width=\textwidth]{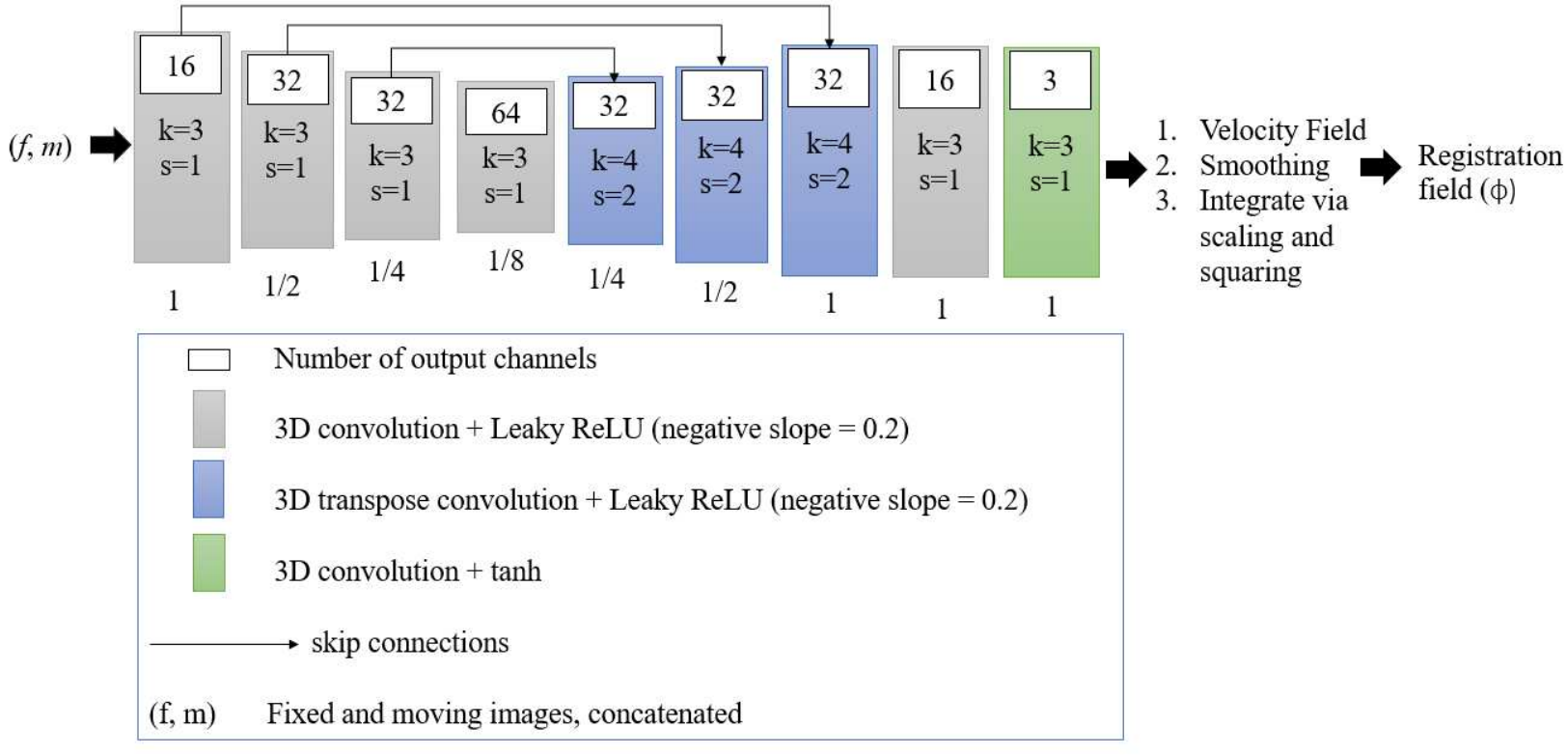}
    \caption{CNN architecture used for registration. The numbers under each convolution representation indicate the volume spatial resolution relative to the input volume. k=kernel size, s=stride.}
    \label{fig:registration_cnn}
\end{figure}

We employ a U-Net-based encoder-decoder architecture. We use blocks of 3D strided convolutions with leaky ReLU activations. We illustrate our architecture in Fig.~\ref{fig:registration_cnn}.

We train a single CNN on all diagnoses, appropriately pairing each subject to its corresponding atlas. We train the VoxelMorph registration network until convergence (81,653 iterations, NVIDIA GeForce RTX 3090 GPU), using a linearly decaying learning rate initialised at $5\times 10^{-4}$ and an Adam optimiser (default $\beta$ parameters, weight decay of $1\times 10^{-5}$). We use Project MONAI spatial and intensity data augmentation\footnote{\label{monai} https://github.com/Project-MONAI/MONAI/.}.

\subsection{CNN segmentation}

3D Attention U-Net is our backbone segmentation architecture, due to its success in segmenting multi-class structures of varying locations and sizes \citep{oktay2018attention}. 

\subsubsection{Segmentation loss function
}
\label{sec:seg_loss}

We use the \textbf{soft dice and cross-entropy loss} ($\textrm{DiceCE}_{loss}$, \cite{hatamizadeh2022unetr}) for all our segmentation experiments, with both cross entropy and dice loss weighted equally.

We additionally investigated network performance using both \textbf{Generalised Dice loss} (GDL, \cite{sudre2017generalised}), and \textbf{GDL+Focal loss} \citep{lin2017focal}. However, we found very poor performance when using both these losses, such as unlearnt small vessels.

Our proposed framework \textit{Attention U-Net LP + Man} is trained using a combined loss

\begin{equation}
\begin{gathered}
\label{eqn:seg_loss}
    \mathcal{L}_{seg} = \lambda _3 \textrm{DiceCE}_{loss}(\textrm{pred}_{multi}, [m_{\textrm{lab}}\cdot \phi ]) \: +  \\
    \lambda _2 \textrm{DiceCE}_{loss}(\textrm{pred}_{joined}, f_{lab})
\end{gathered}
\end{equation}
where $\textrm{pred}_{multi}$ are the multi-class Attention U-Net predictions, $\textrm{pred}_{joined}$ are binary label predictions (multi-class output labels joined together),  $[m_{\textrm{lab}}\cdot \phi]$ are the propagated atlas labels, $f_{lab}$ are the manual binary labels, $\lambda _2$ is the binary loss weight, and $\lambda _3$ the multi-class loss weight. The proposed losses are schematically presented in Fig.~\ref{fig:overall_architecture}.

\subsubsection{Segmentation network implementation details
}
\label{Sec:seg_net_implementation_details}

We use a 3D Attention U-Net \citep{oktay2018attention} implemented in Project MONAI\footnotemark[\getrefnumber{monai}] for automated segmentation, with five encoder-decoder blocks (output channels 32, 64, 128, 256 and 512), convolution and upsampling kernel size of 3, ReLU activation, dropout ratio of 0.5, batch normalisation, and a batch size of 12. We employ an AdamW optimiser with a linearly decaying learning rate, initialised at $1\times10^{-3}$, default $\beta$ parameters and weight decay=$1\times 10^{-5}$. We use intensity and spatial augmentations from Project MONAI\footnotemark[\getrefnumber{monai}], including random intensity shifts, Gaussian smoothing, Gaussian noise, sharpening, contrast adjustments and bias field. 

We train our proposed method (\textit{Attention U-Net LP + Man}, NVIDIA GeForce RTX 3090 GPU) by increasing the binary weight ($\lambda_2$) by 0.05 every 50 epochs until convergence (15809 iterations), with the propagated labels weight set to a fixed value ($\lambda_3$=1). This is to ensure accurate vessel classification while increasing the accuracy of the whole vessels region of interest (ROI).

\subsection{Classification}

We train an anomaly classifier to discern between three classes: CoA, RAA and DAA. We employ DenseNet121 \citep{huang2017densely} as our backbone classifier architecture.

\begin{figure}[h]
     \centering
         \includegraphics[width=0.7\textwidth]{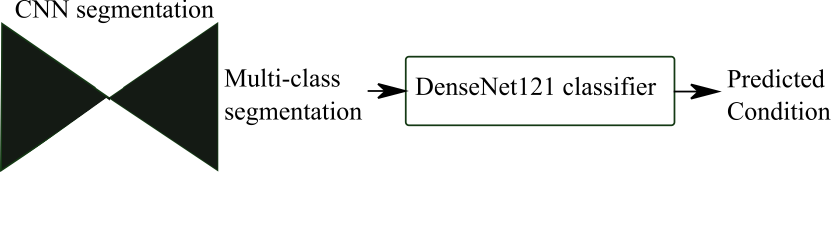}
         \caption{\textbf{DenseMulti}: our classifier experiment trained on the output softmax segmentation prediction. }
         \label{fig:classifier_diagram_seg}
\end{figure}

\textbf{DenseMulti} is our classifier trained jointly with a segmenter network (\textit{Attention U-Net LP + Man}). This is a multi-task framework, where we update the weights of both networks simultaneously and use the features learnt from the segmenter to train the classifier. The inputs to the classifier are the output softmax features from the segmenter network. Fig.~\ref{fig:classifier_diagram_seg} illustrates our joint training framework. We aim to explore whether combined training will encourage the segmenter to learn the characteristic features of each anomaly. 

We additionally explored training a classifier on the latent space of the segmenter network. However, we did not find a significant segmentation improvement.

\subsubsection{Classification loss function}

We employ different optimizers for classifier and segmenter networks and perform backpropagation using the same total loss for both networks. This is expressed as 

\begin{equation}
\label{eq:Ljoint}
    \mathcal{L}_{joint} = \mathcal{L}_{seg} + \lambda_4\mathcal{L}_{class} . 
\end{equation}

$\mathcal{L}_{seg}$ is the weighted sum of DiceCE$_{loss}$ between propagated labels and predictions (multi-class output) and between manual binary labels and predictions (binary output), see Sec.~\ref{sec:seg_loss}.  $\mathcal{L}_{class}$ is the classification cross entropy loss; and $\lambda_4$ is a tunable loss weight. 

\subsubsection{Classification implementation details}

We pre-train our segmentation networks prior to the addition of the classifier (with both propagated and binary labels), as described in Sec.~\ref{Sec:seg_net_implementation_details}. The classifier is also pre-trained (on the softmax segmentation output) keeping the segmenter network weights fixed. This is to ensure that the classifier only learns highly representative features of each anomaly; and that the classifier does not degrade the segmenter network performance. 

We repeat each joint training experiment three times, employing an independently trained segmentation network each time (the same repeated \textit{Attention U-Net LP+Man} experiments described in Sec.~\ref{sec:experiments_explanation_lp_Man}). We use the latest binary weight ($\lambda_2$) from the best pre-trained network (on our validation set) in each case; and tune the classifier weight ($\lambda_4$ in Eq.~\ref{eq:Ljoint}), with the optimal weight being between 6 and 12 for \textbf{DenseMulti}, and 0.1 to 1 for \textbf{DenseBin} and \textbf{DenseImgMulti}, which is a classifier trained only on binary segmentation output (see Sec.~\ref{sec:methods_class_exp}). We find the weight balancing to be very important and tune this parameter for each training experiment (which employs an independently trained segmentation network each time) by observing segmentation loss convergence, classifier validation set scores and visual segmentation inspection on our validation set. We train until reaching the best performance on our validation set (lowest multi-class $\textrm{DiceCE}_{loss}$ on validation set).

\section{Experiments}

\subsection{Training label type ablation study}
\label{sec:experiments_explanation_lp_Man}

As presented in \cite{ramirez2022automated}
we conduct ablation studies on the inclusion of each type of labelling information, e.g. manual binary labels in individual subject images and multi-class labels propagated from the anomaly-specific atlases. We evaluate segmentation performance in the following experiments:

\begin{itemize}
    \item {\textbf{\textit{LP}}: Label Propagation using VoxelMorph.}
    \item {\textbf{\textit{Attention U-Net LP}}: Attention U-Net trained exclusively with propagated labels ($\lambda_2=0$ in Eq.~\ref{eqn:seg_loss})}.
    \item {\textbf{\textit{Attention U-Net Man}}: Attention U-Net trained exclusively with manually segmented binary labels ($\lambda_3 = 0$)}. 
    \item {\textbf{\textit{Attention U-Net LP+Man}}: Attention U-Net trained with both manual binary labels and propagated labels (Eq.~\ref{eqn:seg_loss}).}

    \end{itemize}

Expanding on our previous work \citep{ramirez2022automated}, we repeat our segmentation experiments three times, to account for network stochasticity.

\subsection{Weak supervision}
\label{sec:weak_sup_exp}

We study the impact a varying amount of manually generated labels have on segmentation network performance. In this ablation study, we solely examine the weak binary manual labels as they are meticulously created by expert fetal cardiac clinicians, while the propagated labels are automatically derived.

Six segmentation experiments are compared, with varying percentages of manual binary labels (Man) included: 0\% Man (no manual binary labels in training, i.e. \textit{Attention U-Net LP}), 5\% Man, 25\% Man, 50\% Man, 75\% Man, and 100\% Man (i.e. our proposed \textit{Attention U-Net LP + Man}). These percentages are computed for each anomaly (out of their respective total number of training cases). In all these experiments the propagated labels are fully used. The same implementation details as described in Sec.~\ref{Sec:seg_net_implementation_details} are employed.

\subsection{Classification experiments}
\label{sec:methods_class_exp}

We conduct four main experiments to evaluate our aortic arch anomaly classifiers.

\textbf{DenseMulti} is a DenseNet121 trained from the softmax output of the \textbf{multi-class} segmenter network (\textit{Attention U-Net LP + Man}, see Fig.~\ref{fig:classifier_diagram_seg}).

\textbf{DenseBin} is a DenseNet121 trained from the softmax output of our \textbf{binary} segmenter network (\textit{Attention U-Net Man}). 

\textbf{DenseImage}, is a DenseNet121 trained exclusively on the volume images as input. 

\textbf{DenseImgMulti}, is a DenseNet121 trained on the multi-class softmax output of \textit{Attention U-Net LP + Man}, concatenated with the input T2w image.

We pre-train \textbf{DenseBin}, \textbf{DenseMulti} and \textbf{DenseImgMulti} using the frozen weights of each respective segmenter network (\textit{Attention U-Net Man}, and \textit{Attention U-Net LP + Man}). We report these results under each classifier name followed by \textbf{Separate}. We then train our pre-trained classifier and segmenter networks jointly, which is our multi-task approach, and report results under the title \textbf{Joint}.

We repeat each experiment three times, employing our three independently pre-trained segmenter network rounds.

\subsection{Evaluation metrics}
\label{sec:res_specification}

\subsubsection{Quantitative analysis}
We manually generated multi-class ground truth (GT) labels for our test set (N=40) via ITK-SNAP \citep{py06nimg}. We report similarity metrics including multi-class Dice scores, recall, precision, average surface distance (ASD), and 95th percentile of the Hausdorff Distance (HD95).  We repeat our segmentation experiments three times and average results. We also include a short analysis studying the impact of image quality on performance. 

\subsubsection{Qualitative analysis}
\label{sec:qualitative_analysis_definition}
There are inherent limitations to conducting a quantitative analysis exclusively, particularly given the quality of our data and the task at hand. We find the topological correctness of the anomaly area, our key segmentation objective, to not be reflected in our quantitative metrics described above. We thereby devise a \textbf{qualitative analysis} strategy, which involves manual inspection and scoring of the anomaly area of each test set subject. 

We assess the topological correctness of the aortic arch (anomaly area), as described in Table~\ref{tab:score_summary}, where a score of 1 (best) represents an aortic arch with topologically correct anomaly delineation, and a score of 3 (worst) indicates topologically incorrect delineation. Types of topological errors for segmentations with a score of 3 include split aortic arch, indiscernible aortic arch (merged), an unsegmented or partially segmented right arch in a DAA case (split arch), a segmented left arch for RAA cases (double arch segmented), and a segmented right arch for CoA cases (double arch segmented). The term oversegmentation refers to erroneously labelling background voxels as an anatomical structure, while rarely labelling structure as background.  This typically results in erroneously thick vessels.

\begin{table}[h]
\centering
\caption{Description of our anomaly area topological assessment. HN refers to head and neck vessels, and AD arterial duct.}
\begin{tabular}{cc}
\hline
\multicolumn{2}{c}{\textbf{Aortic arch score}}        \\ \hline
1 & Topologically correct aortic arch    \\

2 & Oversegmented aortic arch merging into HN vessels or AD.  \\

3  & Topologically incorrect aortic arch

\end{tabular}
\label{tab:score_summary}
\end{table}

\section{Results}

\subsection{Training label type ablation study}

Here we present metrics comparing our proposed segmentation framework which learns from both binary individual labels and multi-label atlases, against the inclusion of just one type of labelling (see Sec.~\ref{sec:experiments_explanation_lp_Man}). Differences to \cite{ramirez2022automated} 
are due to adaptations made to the atlas in pulmonary arteries and exclusion of pulmonary veins (following discussions with clinicians), and stochasticity due to network retraining. All metrics displayed are on unseen test set, comparing to our manually generated GT.

\subsubsection{Whole vessel ROI}

\begin{figure}[ht]
    \centering
    \includegraphics[width=0.9\textwidth]
    {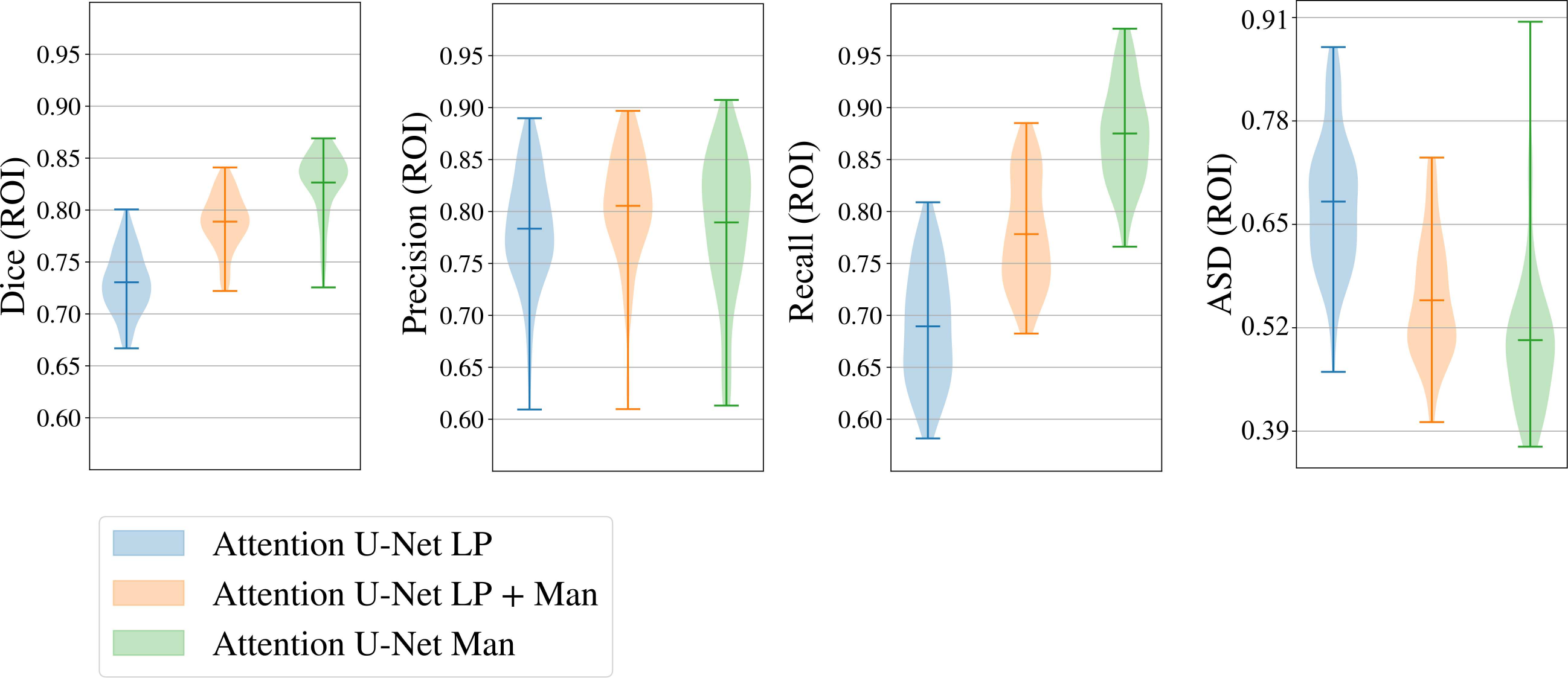}
    \caption{Quantitative ROI metrics for our labelling ablation experiments. The horizontal lines represent the means.}
    \label{fig:roi_scores_man_lp}

\end{figure}

Here we present metrics for the whole vessels' ROI in violin plots in Fig.~\ref{fig:roi_scores_man_lp}. In the case of multi-label segmentation, the vessels were joined into a single region prior to the metrics calculations.

\textit{ Attention U-Net Man}, which is the network trained on binary labels, displays the highest mean vessels' ROI dice ($0.83 \pm 0.03$). However, further analysis reveals a much higher average recall ($0.87 \pm 0.05$) than precision ($0.79 \pm 0.07$), which points to significant oversegmentation. This is clearly visible in Fig.~\ref{fig:merge_and_hn_man_lpman}. We can observe that oversegmentation results in the merging of the neighbouring vessels, obstructing the visualisation of the fetal cardiac vessels' anatomy.

Conversely, the proposed network \textit{Attention U-Net LP + Man} has a more balanced performance with a mean Dice $=0.79 \pm 0.03$, a mean precision $=0.80 \pm 0.05$, and a mean recall $=0.78 \pm 0.05$. The resulting segmentations (Fig.~\ref{fig:merge_and_hn_man_lpman}) offer a clear depiction of individual vessels, which is required for our clinical application.

\textit{Attention U-Net Man} also has the lowest mean ASD of $0.51 \pm$ 0.091 mm, which increases to $0.55 \pm 0.08$ for the proposed \textit{Attention U-Net LP + Man}. However, \textit{Attention U-Net Man} presents outliers with ASD even greater than \textit{LP}, whereas \textit{Attention U-Net LP + Man} is much more consistent across cases.

The network trained on propagated labels only \textit{Attention U-Net LP} has the lowest performance in all the metrics.

\subsubsection{Individual fetal cardiac vessels}
\label{res:indiv_vess}

Table \ref{tab:multi-results-lp-Man} displays metrics comparing our multi-class experiments. We compare our full approach (\textit{Attention U-Net LP + Man}) to Attention U-Net trained exclusively on propagated labels (\textit{Attention U-Net LP}) and VoxelMorph label propagation (\textit{LP}). These results indicate that adding the binary labels to the training of the segmentation network offers a significant improvement for Dice and ASD metrics for all vessels. Further, Attention U-Net trained on propagated labels improves on label propagation alone.

\begin{table}[h]
\caption{Similarity metrics between our multi-class experiment predictions and our manually generated GT. We highlight the best scores in bold. HN indicates the head and neck vessels (LSA, LCCA, BCA, RSA). The three bottom rows report metrics averaged over all vessels. We highlight with an * the vessels with a p-value$<0.05$ (Mann-Whitney U test) (non-parametric) compared to \textit{Attention U-Net LP + Man}. } 
\begin{tabular}{l|cc|cc|cc}
\textbf{} & \multicolumn{2}{c|}{\textit{\textbf{LP}}} & \multicolumn{2}{c|}{\textit{\textbf{Attention U-Net LP}}} & \multicolumn{2}{c}{\textit{\textbf{\shortstack{Attention U-Net \\ LP +   Man}}}} \\ \hline
Vessel    & Dice                & ASD                 & Dice                        & ASD                         & Dice                            & ASD                            \\ \hline
SVC       & 0.65 (0.15)*         & 0.88 (0.60)*         & 0.70 (0.06)*                 & 0.70 (0.20)*                 & \textbf{0.76 (0.06)}            & \textbf{0.56 (0.16)}           \\
LPA       & 0.58 (0.15)*         & 0.88 (0.93)*         & 0.63 (0.08)*                 & 0.72 (0.30)*                & \textbf{0.65 (0.08)}            & \textbf{0.68 (0.29)}           \\
RPA       & 0.55 (0.12)*         & 0.80 (0.44)*         & 0.61 (0.07)*                 & 0.71 (0.23)                 & \textbf{0.63 (0.07)}            & \textbf{0.67 (0.21)}           \\
Aortic arch     & 0.58 (0.13)*         & 0.87 (0.46)*         & 0.62 (0.06)*                 & 0.79 (0.16)*                 & \textbf{0.76 (0.04)}            & \textbf{0.54 (0.08)}           \\
AD        & 0.70 (0.16)*         & 0.66 (0.57)*         & 0.76 (0.07)*                 & 0.49 (0.11)*                 & \textbf{0.78 (0.07)}            & \textbf{0.44 (0.12)}           \\
DAO       & 0.77 (0.11)*         & 0.63 (0.40)*         & 0.81 (0.04)*                & 0.52 (0.20)*                 & \textbf{0.84 (0.04)}            & \textbf{0.43 (0.18)}           \\
MPA       & 0.74 (0.11)*         & 0.71 (0.25)*         & 0.76 (0.06)*                 & 0.68 (0.16)*                 & \textbf{0.81 (0.05)}            & \textbf{0.55 (0.13)}           \\
HN        & 0.31 (0.21)*         & 1.50 (1.27)*         & 0.38 (0.20)                 & 1.24 (0.94)*                 & \textbf{0.40 (0.19)}            & \textbf{1.20 (0.93)}           \\ \hline
Avg. Dice & \multicolumn{2}{c|}{0.54 (0.24)}          & \multicolumn{2}{c|}{0.59 (0.20)}                          & \multicolumn{2}{c}{\textbf{0.63 (0.21)}}                         \\
Avg. HD95 & \multicolumn{2}{c|}{3.14 (1.07)}          & \multicolumn{2}{c|}{2.71 (1.81)}                          & \multicolumn{2}{c}{\textbf{2.50 (1.87)}}                         \\
Avg. ASD  & \multicolumn{2}{c|}{2.26 (0.97)}          & \multicolumn{2}{c|}{0.85 (0.63)}                          & \multicolumn{2}{c}{ \textbf{0.77 (0.63)}}                        
\end{tabular}
\label{tab:multi-results-lp-Man}
\end{table}

\subsubsection{Visual inspection}

\begin{figure}[h]
    \centering
    \includegraphics[width=0.8\textwidth]{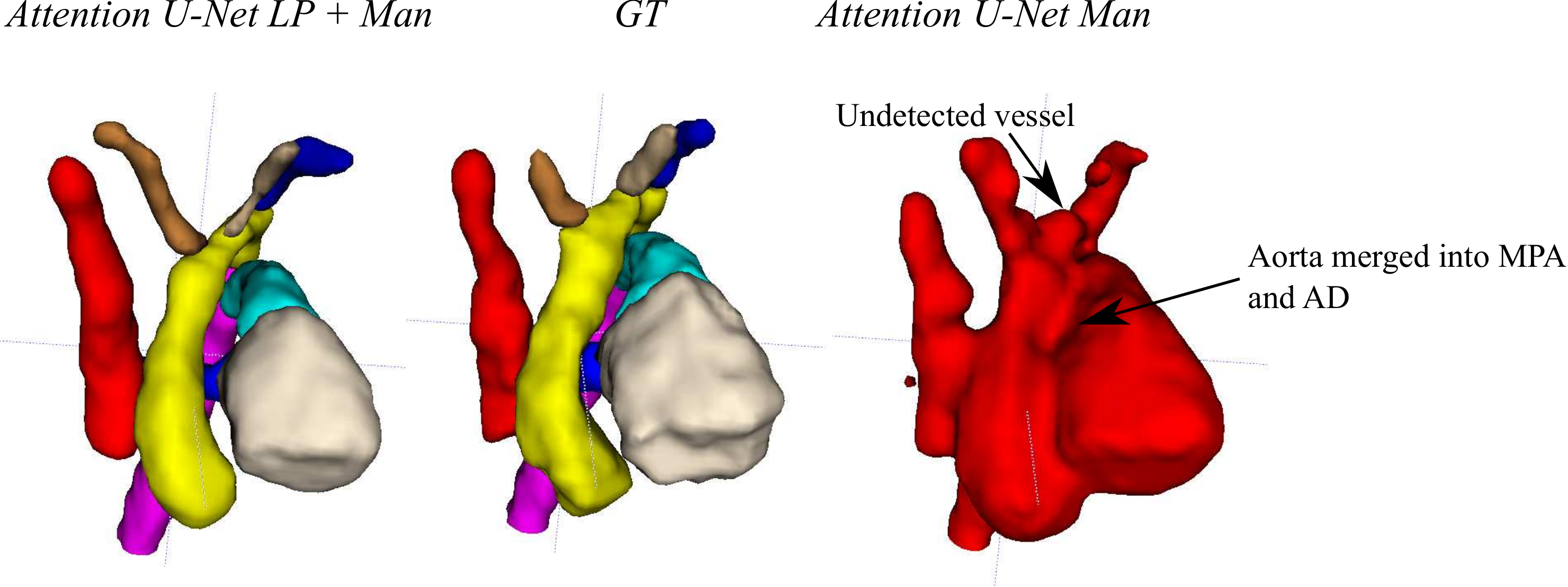}
    \caption{Depiction of aorta topological inaccuracies and undetected small vessels commonly occurring in \textit{Attention U-Net Man}.}
    \label{fig:merge_and_hn_man_lpman}
\end{figure}

Nevertheless, we find these quantitative metrics to not be fully descriptive of the anatomical correctness of the segmentation, a decisive feature when segmenting anomalies. Following visual inspection, we can conclude two main advantages of our multi-class approach: \textbf{\textit{Attention U-Net LP + Man} is more consistent in both small vessel detection and topological correctness of the anomaly area compared to \textit{Attention U-Net Man}}. The latter is quantified in Sec.~\ref{sec:results_qualitative}. Fig.~\ref{fig:merge_and_hn_man_lpman} includes a depiction of these two prominent issues observed across \textit{Attention U-Net Man} experiments: undetected small vessels, and merging of the aorta into surrounding vessels. 

\subsubsection{Visualisation of the latent representation}

\begin{figure}[h]
    \centering
    \includegraphics[width=0.9\textwidth]{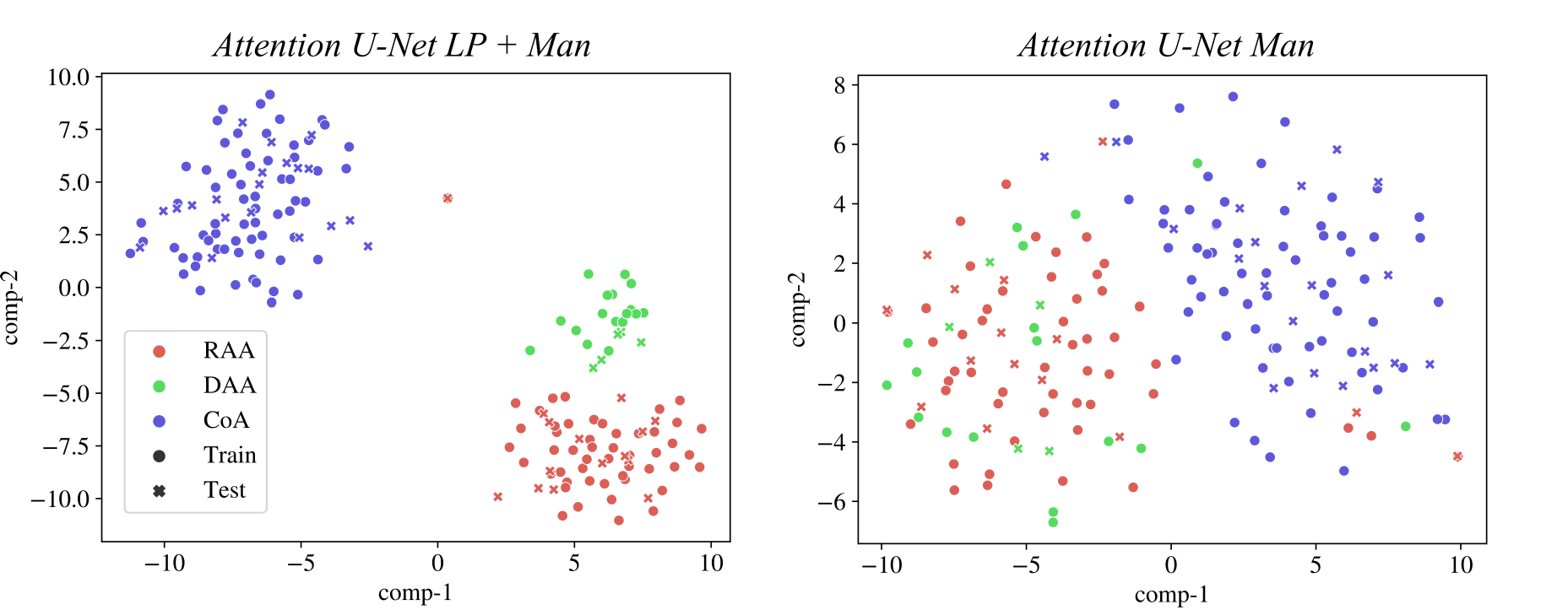}
    \caption{t-SNE bottleneck reduced features for one training round of our combined label framework (multi-class, \textit{Attention U-Net LP + Man}) and our binary network (\textit{Attention U-Net Man}).}
    \label{fig:tsne_man_vs_lp_man}
\end{figure}

We inspect the anomaly-specific discernment of our labelling ablation networks by visualising t-SNE reduced latent space features (Fig.~\ref{fig:tsne_man_vs_lp_man}). With this, we can easily examine the effect multi-class and binary labels have on anomaly distinction for our networks. 

We can conclude that our multi-class network \textit{Attention U-Net LP + Man} offers highly clustered groups for anomalies even in the bottleneck features, contrarily to our binary network (\textit{Attention U-Net Man}) which especially struggles to discern between RAA and DAA cases, as supported by our qualitative analysis (Sec.~\ref{sec:results_qualitative}). We include further t-SNE plots comparing our various experiments in Appendix~\ref{appendix:latent}. 

\subsubsection{Weak supervision ablation}

\begin{figure}[h]
     \centering
     \begin{subfigure}[b]{0.49\textwidth}
         \centering
         \includegraphics[width=\textwidth]{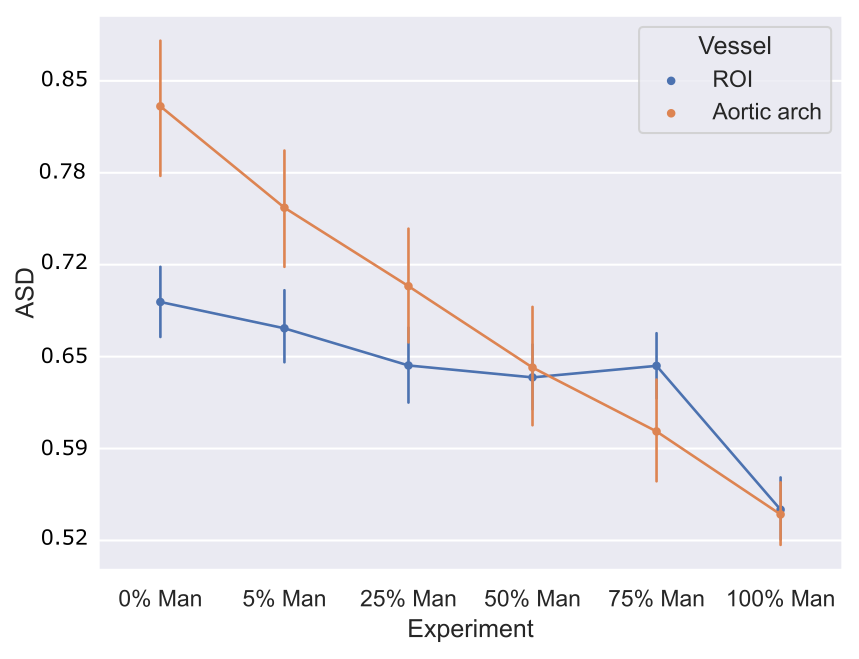}
         \caption{ASD scores (lower = better).}
         \label{fig:asd_weakbin}
     \end{subfigure}
     \hfill
     \begin{subfigure}[b]{0.49\textwidth}
         \centering
         \includegraphics[width=\textwidth]{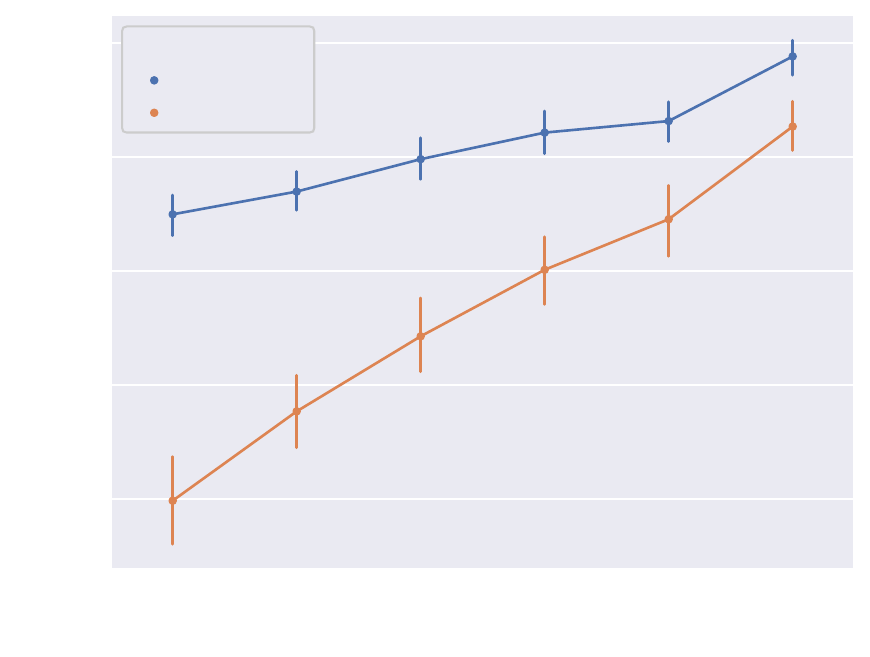}
         \caption{Dice scores (higher = better).}
         \label{fig:dice_weakbin}
     \end{subfigure}

        \caption{Test set metrics for the vessels' ROI, and aortic arch predicted label comparing the impact of adding binary manual labels (Man) during training. Error bars represent the 95\% confidence interval, computed via bootstrapping (non-parametric uncertainty representation). }
        \label{fig:weakbin}
\end{figure}

Fig.~\ref{fig:weakbin} includes Dice and ASD scores for the vessels' ROI and aortic arch label, for an increasing number of training cases with manually generated binary labels. Our findings indicate that including binary labels in training results in a statistically significant improvement in segmentation performance (p-value $<$ 0.05, Mann-Whitney U test) compared to not including any binary labels (0\%, \textit{Attention U-Net LP}) when 25\% or more of the manual labels were included. For 5\% of included manual labels, the improvement in Dice and ASD is only significant for the aortic arch region, not the whole ROI.

\subsection{Classification of anomalies}

\subsubsection{Classification performance}

Table~\ref{tab:classifier_scores} presents accuracy and balanced accuracy scores (average recall obtained on each class) for each classifier experiment, as well as recall scores for each individual anomaly. In addition to comparing different types of features for classification (multi-label segmentation, binary segmentation and original image), we also compare separate and joint training of segmenter and classifier networks. The confusion matrices are provided in Appendix~\ref{appendix:condition_classifier}.

We observe that while joint training of the classifier with binary segmentation improves the classification results, this is not the case for multi-class segmentation. For multi-class segmentation, the differences in classification performance in jointly and separately trained frameworks are caused by a small number of subjects, with three misclassifications (two individual subjects) over the three training rounds in the joint framework (one misclassification per round), and two misclassifications in a single subject 
in the separate framework. There is therefore no clear improvement with joint training in the multi-label case, however, the classification performance is high, with an accuracy of 0.97 for \textbf{DenseMulti} and 0.99 for \textbf{DenseImgMulti}.

In general, CoA cases are always correctly classified, while the misclassifications happen between RAA and DAA cases.

Overall, the anomaly classification from multi-label segmentation clearly outperforms the classification from binary segmentations or directly from the images. However, when incorporating image information concatenated with the output segmentation, the classifier achieves the highest performance. This image data inclusion allows the network to focus on global features and becomes less reliant on topologically accurate segmentation predictions compared to \textbf{DenseMulti}. Consequently, \textbf{DenseImgMulti}'s high accuracy is not truly reflective of segmentation performance, as correct classification can still occur even with important topological errors in the anomaly area.

\begin{table}[h]
\caption{Accuracy and recall scores for our classifier experiments, averaged over all three training rounds. Separate = prior to joint segmenter and classifier training, Joint = after joint training. Bl. Acc. = balanced accuracy. 
}
\label{tab:classifier_scores}
\begin{tabular}{l|ccccc}
                       & \textbf{Accuracy}  & \textbf{Bl. Acc.} & \textbf{Recall CoA} & \textbf{Recall RAA} & \textbf{Recall DAA} \\ \hline
\textbf{DenseMulti}    &                    &                       &                     &                     &                     \\
Separate               & 0.98 (0.01)        & 0.95 (0.04)           & \textbf{1.0 (0.0)}  & \textbf{1.0 (0.0)}  & 0.87 (0.11)         \\
Joint                  & 0.97 (0.0)         & 0.98 (0.0)            & \textbf{1.0 (0.0)}  & 0.93 (0.0)          & \textbf{1.0 (0.0)}  \\
\textbf{DenseBin}      &                    &                       &                     &                     &                     \\
Separate               & 0.88 (0.01)        & 0.90 (0.01)           & 0.95 (0.0)          & 0.75 (0.04)         & \textbf{1.0 (0.0)}  \\
Joint                  & 0.92 (0.02)        & 0.91 (0.05)           & 0.97 (0.03)         & 0.89 (0.10)         & 0.81 (0.17)         \\
\textbf{DenseImgMulti} &                    &                       &                     &                     &                     \\
Separate               & \textbf{1.0 (0.0)} & \textbf{1.0 (0.0)}    & \textbf{1.0 (0.0)}  & \textbf{1.0 (0.0)}  & \textbf{1.0 (0.0)}  \\
Joint                  & 0.99 (0.01)        & 0.99 (0.01)           & \textbf{1 (0.0)}    & 0.98 (0.04)         & \textbf{1.0 (0.0)}  \\
\textbf{DenseImage}    & 0.82 (0.07)        & 0.79 (0.09)           & 0.98 (0.03)         & 0.64 (0.01)         & 0.62 (0.30)        
\end{tabular}
\end{table}

\begin{figure}[h]
    \centering
    \includegraphics[width=\textwidth]{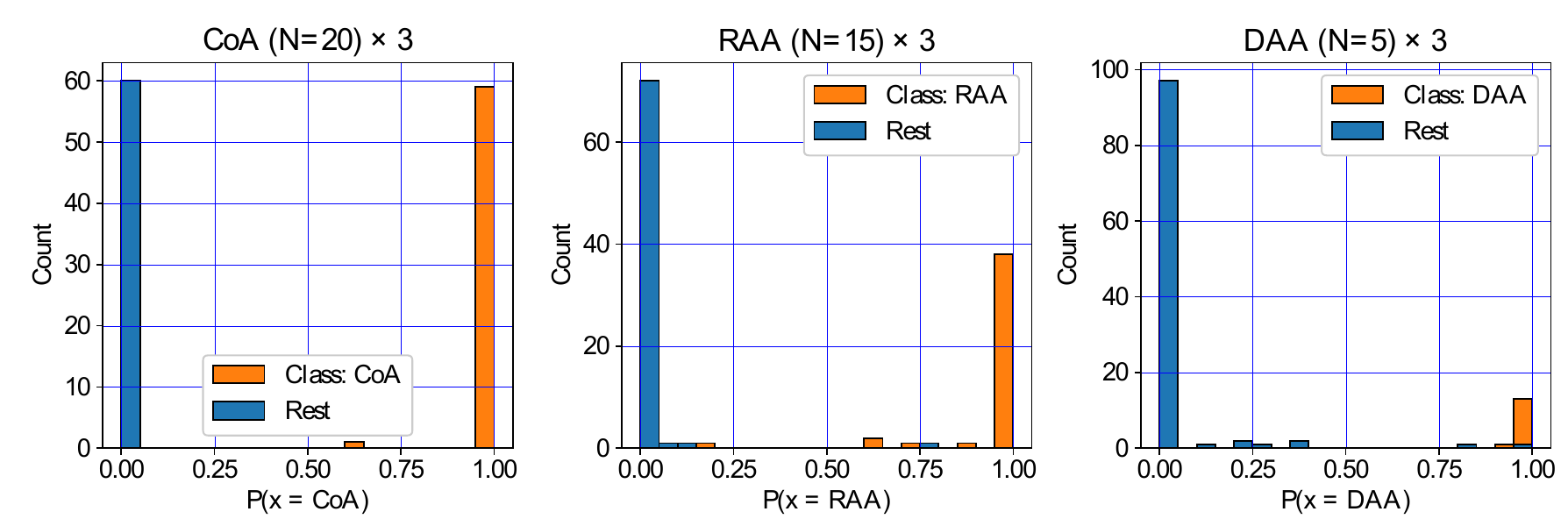}
    \caption{Count of probability distributions for all three rounds of \textbf{DenseMulti} classifier experiment on our test set.}
    \label{fig:proba_class}
\end{figure}

We display our \textbf{DenseMulti} class probabilities in Fig.~\ref{fig:proba_class} for each anomaly (one vs. rest). These include the results of our three training rounds. We observe a higher number of cases with a lower softmax probability for RAA and DAA groups, meaning that our classifier is less confident when distinguishing between RAA and DAA and the rest of our groups, however very confident for CoA predictions. This is understandable, given the anatomical similarities between RAA and DAA subjects (Fig.~\ref{fig:condition_desc}).

\subsubsection{Analysis of misclassified cases}
\label{sec:misclassifications}

\begin{figure}[h!]
    \centering
    \includegraphics[width=0.9\textwidth]{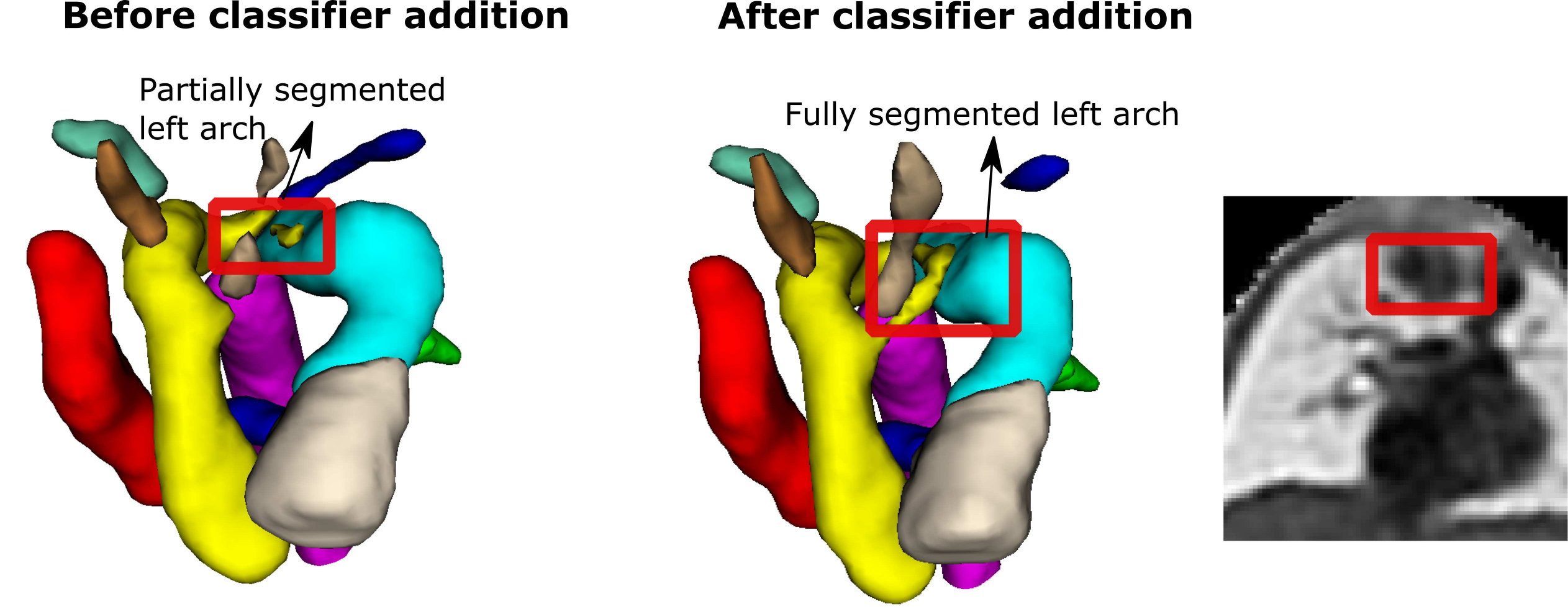}
    \caption{Arrow signals the double arch erroneously segmented in an RAA subject, resulting in misclassification. We highlight the blurrier region where this error occurs in the image with a rectangle. }
    \label{fig:da_seg_class}
\end{figure}

Here we provide further detail on our \textbf{DenseMulti} and \textbf{DenseImgMulti} misclassifications. We obtain one misclassified case per round with \textbf{DenseMulti}, with one case repeated across two of our training rounds. This same case is the one misclassification for \textbf{DenseImgMulti}. This case is an RAA subject erroneously classified as DAA due to the double arch being wrongly segmented (Fig.~\ref{fig:da_seg_class}). This particular case presented lower visibility in the anomaly area, with more than one independently trained network partially segmenting (erroneously) the left arch. We highlight the blurriness in this region causing the oversegmentation. This particular case already presented topological issues prior to adding the classifier (Fig.~\ref{fig:da_seg_class}, left). The addition of the classifier clearly tries to erroneously correct this initial error, by fully segmenting the double arch, i.e. by pushing the segmentation prediction closer to one of the aortic arch anomalies. Importantly, this case presents topological issues for one of our joint \textbf{DenseImgMulti} experiments but was correctly classified. These findings suggest \textbf{DenseImgMulti} to be a more reliable classifier when the segmentation is faulty, as it incorporates image information. Consequently, it cannot be utilized as a quality control tool to identify erroneous segmentations.

\begin{figure}[h!]
    \centering
    \includegraphics[width=0.9\textwidth]{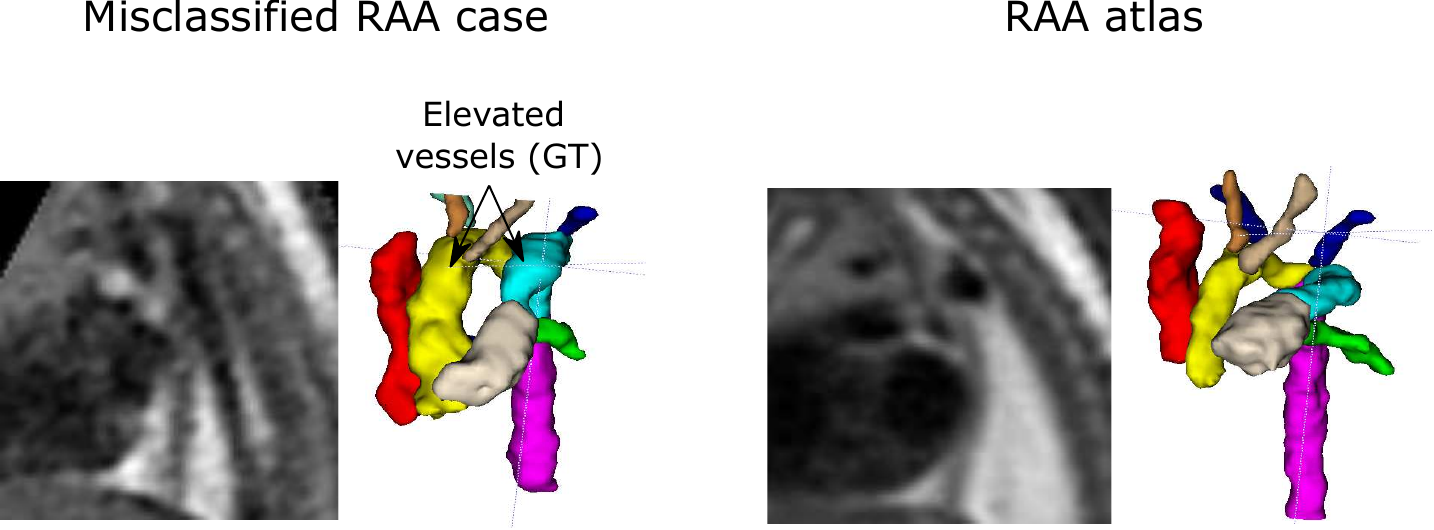}
    \caption{Mislabelled RAA as DAA in one of our rounds due to low image quality and misalignments to the atlas, with a highly elevated aorta and AD compared to the atlas.}
    \label{fig:misclassed2}
\end{figure}

Our second misclassification is another RAA case mislabelled as DAA due to poor image quality and misalignment to the atlas (Fig.~\ref{fig:misclassed2}). Our segmentation here is accurate, however, the aortic arch and AD are highly elevated compared to the atlas (Fig.~\ref{fig:misclassed2}). This may have been misclassified after joint training due to slight overfitting to the training data. However, it is only one case (out of 120 in total).

\subsubsection{Segmentation improvements due to anomaly classifier }
\label{subsec:classifier_improv}

\begin{figure}[h]
    \centering
    \includegraphics[width=\textwidth]{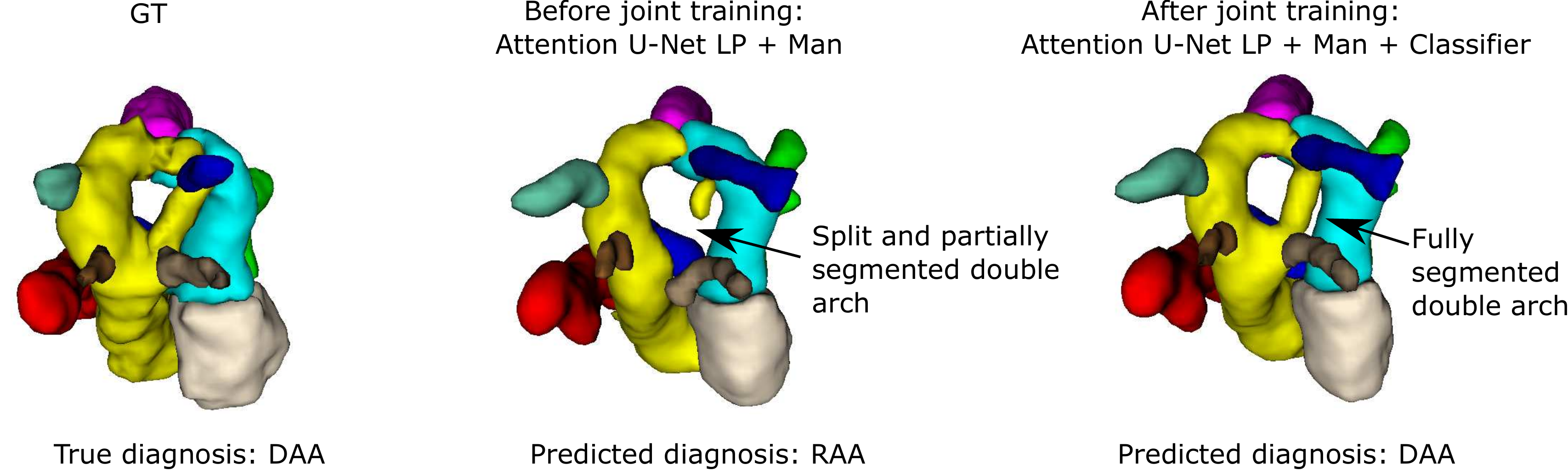}
    \caption{Test case where we find a significant topological improvement of the double arch, resulting in a corrected anomaly prediction. }
    \label{fig:classifier_improv_segclass}
\end{figure}

\begin{figure}[h]
    \centering
    \includegraphics[width=0.55\textwidth]{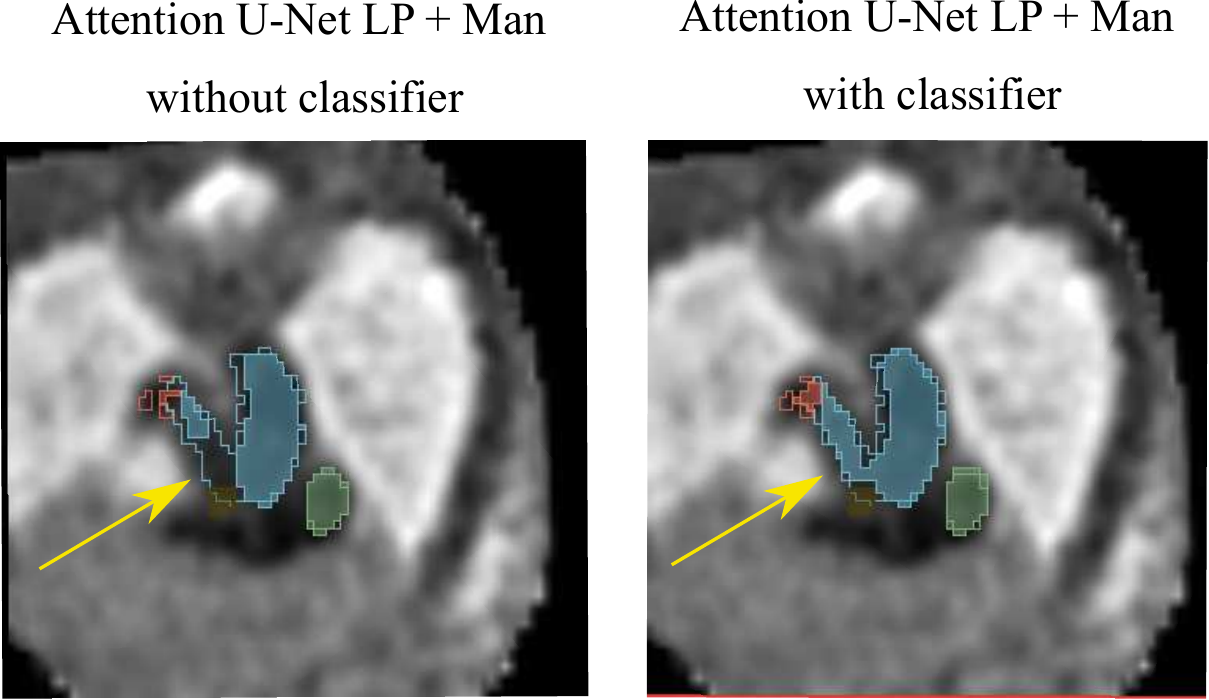}
    \caption{Arrow signals the double arch in a DAA subject, a large portion unsegmented prior addition of the classifier network. The GT is overlaid without filling, i.e. just the contours, while the predictions are filled.}
    \label{fig:classifier_da_improv}
\end{figure}

Regarding segmentation network gains, we find meaningful improvements in our multi-class framework, particularly regarding the completeness of double arch segmentation (Figs.~\ref{fig:classifier_da_improv} and \ref{fig:classifier_improv_segclass}). Although our DAA test set comprises only five cases, we repeat this experiment three times, with independently trained segmenter networks, and find that in \textbf{all correctly classified DAA subjects where the right arch was originally partially segmented or unsegmented, adding a classifier (\textbf{DenseMulti} and DenseImgMulti}) resolved this. The double arch vessel thickness is also improved. This constitutes \textbf{eight subjects with a notable improvement over the three rounds}. We report topological improvements in our overall qualitative analysis, Sec.~\ref{sec:results_qualitative}, Fig.~\ref{fig:qualitative_results}. 

Fig.~\ref{fig:classifier_improv_segclass} depicts a clear example of improved segmentation performance via our multi-task framework, 
consequently correcting the predicted anomaly class.  

We find marginal improvements in aorta metrics for DAA cases (see Tab.~\ref{tab:scores_aorta_classifier}). The fact that these improvements are marginal is likely due to the small surface area the anomaly location comprises, despite being key for robust segmentation.  

\begin{table}[h]
\caption{Marginal improvements in Dice and ASD metrics of the aorta, observed after our joint training. These contain our averaged results for all three training rounds. We highlight in bold our best metrics (higher=better for dice, lower=better for ASD).}
\label{tab:scores_aorta_classifier}
\begin{tabular}{ccc|cc}
    & \multicolumn{2}{c|}{Attention U-Net LP +   Man} & \multicolumn{2}{c}{Attention U-Net LP +   Man + Classifier} \\ \hline
    & Aorta Dice         & Aorta ASD                   & Aorta Dice                   & Aorta ASD                     \\ \hline
CoA & 0.74 (0.04)        & 0.82 (0.15)                & 0.74 (0.04)                  & \textbf{0.81 (0.14)}         \\
RAA & 0.77 (0.02)        & \textbf{0.83 (0.10)}       & 0.77 (0.03)                  & 0.85 (0.11)                  \\
DAA & 0.76 (0.02)        & 0.91 (0.09)                & \textbf{0.77 (0.02)}         & \textbf{0.87 (0.07)}        
\end{tabular}
\end{table}

\begin{figure}[h]
    \centering
    \includegraphics[width=0.85\textwidth]{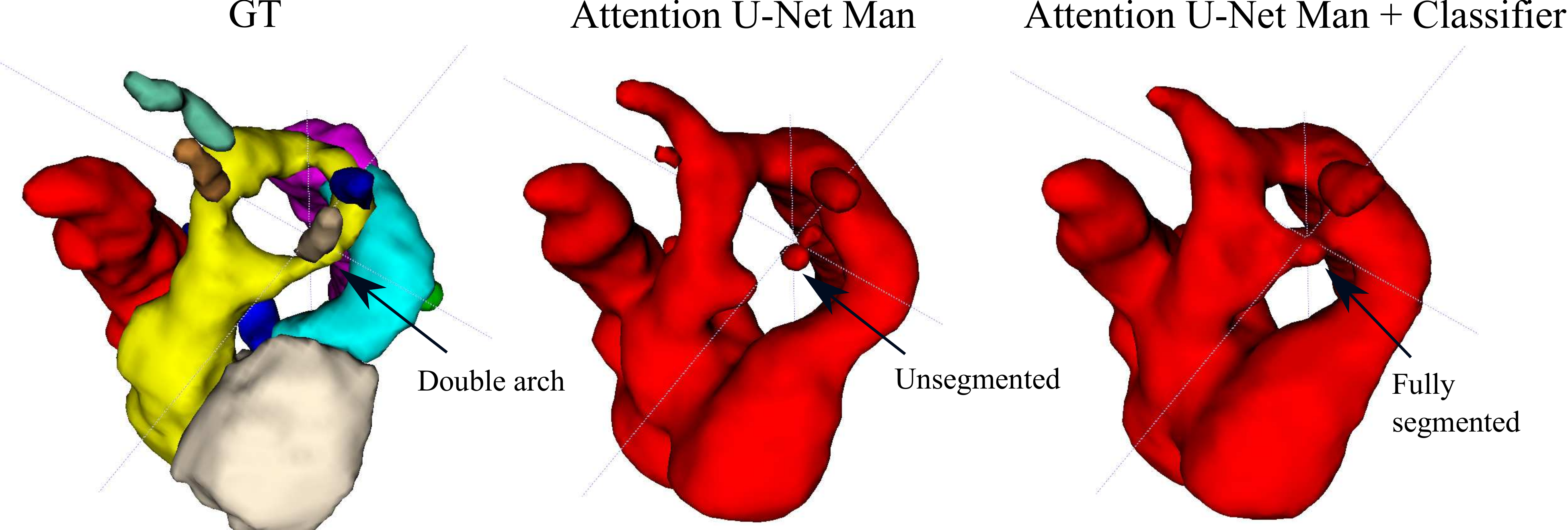}
    \caption{Improvements observed in double arch segmentation after adding a classifier (\textbf{DenseBin}) to our binary framework.}
    \label{fig:densebin_improv}
\end{figure}

Regarding our binary framework (\textit{Attention U-Net Man}), we find that adding a classifier (\textbf{DenseBin}) affects the segmentation output with both improvements observed in double arch segmentation (Fig.~\ref{fig:densebin_improv}) and degradation in certain cases such as heightened vessel merging. 

We postulate that this difference in performance is due to both (a) the anomaly classifier being presented with a more challenging task (harder to identify aorta and anomaly area from a single vessel), and (b) labelling variations due to inter-observer variability, contrasting with our propagated atlas labels.

\subsection{Image quality analysis}

In order to study the impact image quality has on our predictions, a concise analysis of image quality against quantitative network performance is included here. 

Two factors are considered: the reconstruction process (SVR \cite{kuklisova2012reconstruction, kainz2015fast} or DSVR \cite{uus2020deformable}), and a manual image quality assessment. The latter consists of manually revising and scoring the test set images from 1-3 (best to worst), based on image noise, visibility, and vessel sharpness, with 1 being the highest quality, 2 being average quality, and 3 being poor quality and visibility.

A note of consideration is that our test set is majoritarily formed of SVR reconstructions (lower quality, N=102 across all three repeated training rounds), with only 18 DSVR cases. 

\begin{figure}[h]
     \centering
     \begin{subfigure}[b]{0.49\textwidth}
         \centering
         \includegraphics[width=\textwidth]{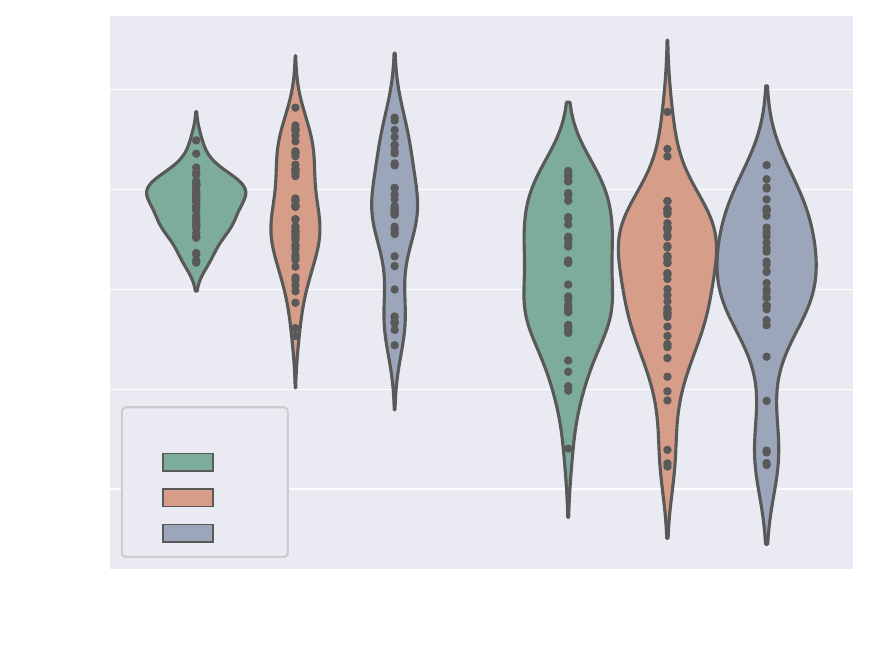}
         \caption{Dice scores.}
         \label{fig:iqs_dice}
     \end{subfigure}
     \hfill
     \begin{subfigure}[b]{0.49\textwidth}
         \centering
         \includegraphics[width=\textwidth]{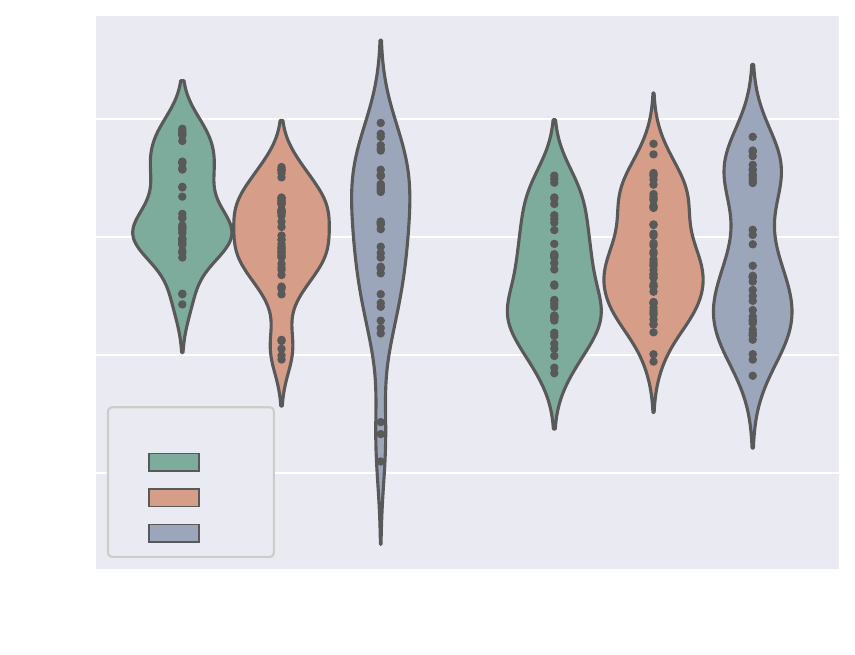}
         \caption{Precision and sensitivity.}
         \label{fig:iqs_prec}
     \end{subfigure}

        \caption{Test set metrics comparing the impact image quality (1$\rightarrow$3 = best$\rightarrow$ worst) has on network performance. }
        \label{fig:iqa}
\end{figure}

Fig.~\ref{fig:iqs_prec} includes precision and sensitivity scores of the vessels' ROI comparing performance for our manual image quality assessment, with Fig.~\ref{fig:iqs_dice} depicting Dice scores for the vessels' ROI and aortic arch region. We generally observe a higher number of lower-scoring outliers in lower-quality images (score = 3), however, we find differences between groups to be statistically non-significant.

\begin{figure}[h]
     \centering
     \begin{subfigure}[b]{0.49\textwidth}
         \centering
         \includegraphics[width=\textwidth]{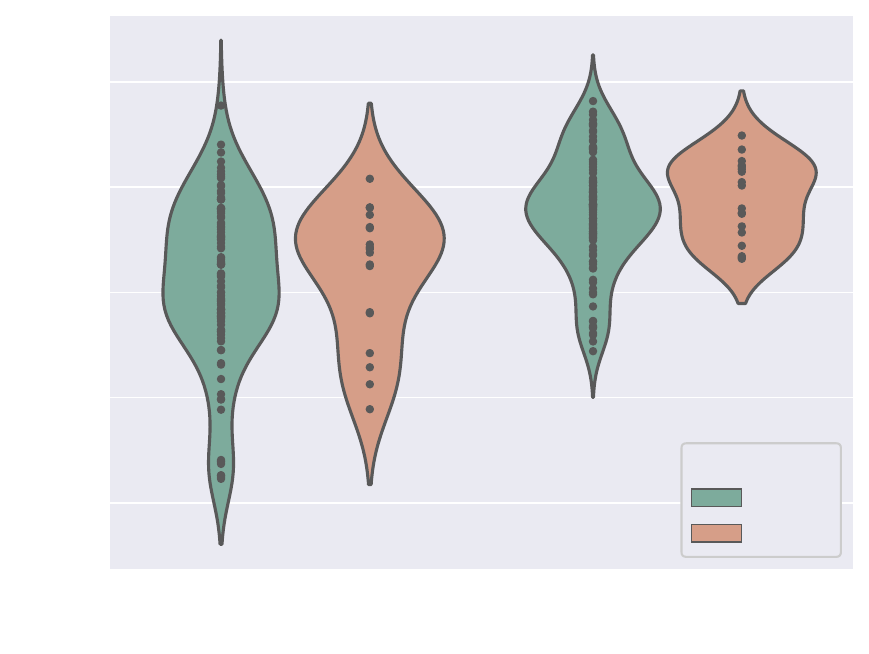}
         \caption{Dice scores.}
         \label{fig:dice_dsvr}
     \end{subfigure}
     \hfill
     \begin{subfigure}[b]{0.49\textwidth}
         \centering
         \includegraphics[width=\textwidth]{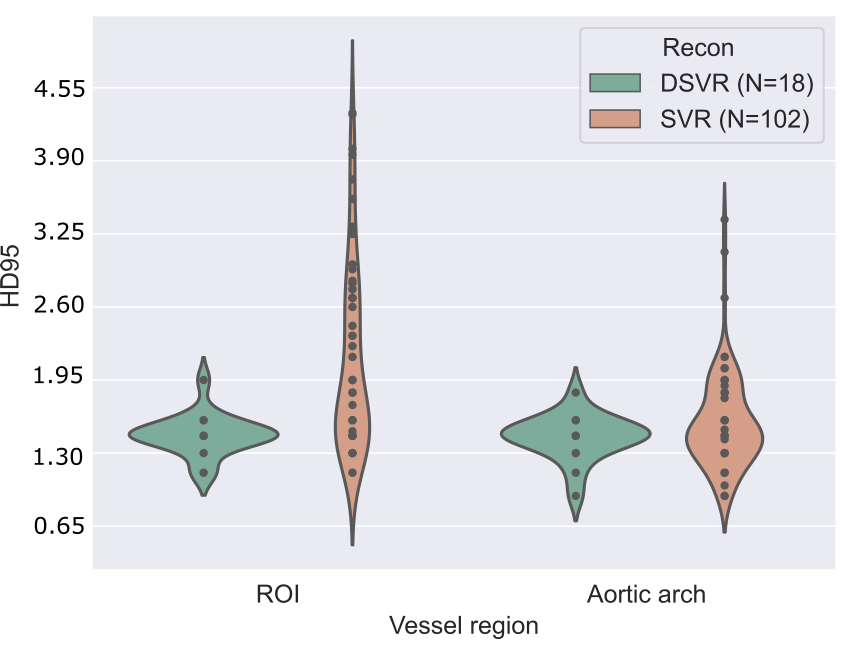}
    \caption{HD95 (lower=better) scores.}
    \label{fig:dsvr_svr}
     \end{subfigure}

        \caption{Test set metrics comparing performance for DSVR and SVR reconstructions.  }
        \label{fig:recon}
\end{figure}

Regarding performance on image reconstruction type (Fig.~\ref{fig:recon}), we find a statistically significant difference (p-value = $3.97\times10^{-4}$ for a Mann-Whitney U test) in performance in HD95 of the vessels' ROI (Fig.~\ref{fig:dsvr_svr}), with DSVR cases presenting lower scores (better performance).

Concerning classifier performance, we find all misclassifications from \textbf{DenseMulti} and \textbf{DenseImgMulti} to be exclusively SVR reconstructions. 

\subsection{Qualitative results}
\label{sec:results_qualitative}

\begin{figure}[h]
    \centering
    \includegraphics[width=\textwidth]{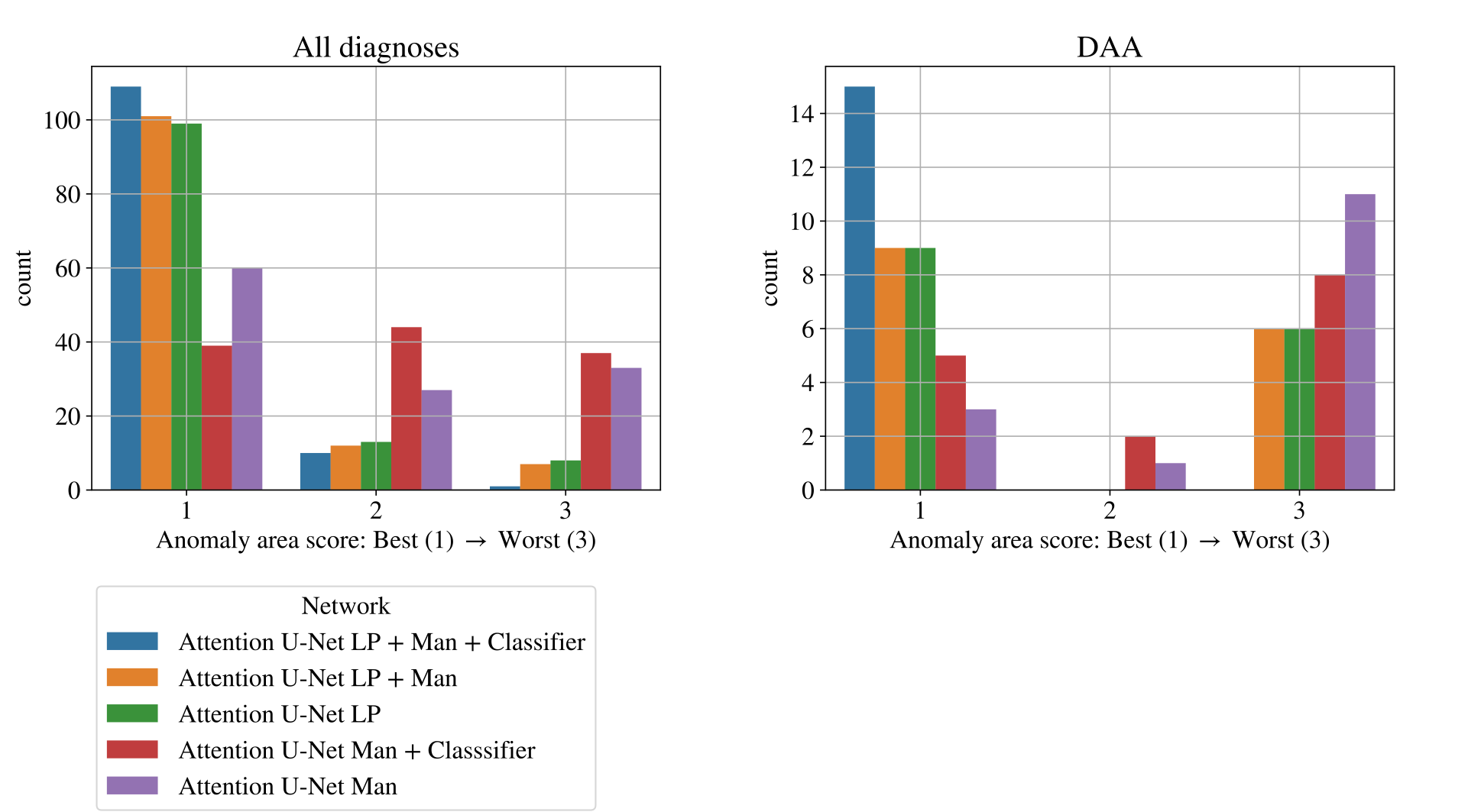}
    \caption{Qualitative analysis results on topology correctness of the anomaly area (aortic arch) on our test set for \textbf{all diagnoses} (left), and for DAA cases only (right). Each training experiment was repeated three times, and we include all of our results here (i.e. we have three sets of results for each of our 40 test set cases, one for each round, so the total number of values for each technique is 120)}. 
    \label{fig:qualitative_results}
\end{figure}

We find that the quantitative metrics presented in previous sections do not always capture the topological and anatomical correctness of the anomaly area, which is key for our clinical application. Accordingly, we present a qualitative topological anomaly area analysis for our segmentation predictions (Sec.~\ref{sec:res_specification} Tab.~\ref{tab:score_summary}). Fig.~\ref{fig:qualitative_results} compares our main ablation studies for classifier inclusion and labelling information for all diagnoses, with our full framework being \textbf{\textit{Attention U-Net LP + Man + Classifier}}. We include the performance on DAA test set exclusively in Fig.~\ref{fig:qualitative_results} left. See Appendix~\ref{appendix:qualitative_extended} for similar CoA and RAA plots.

Figs.~\ref{fig:qualitative_results} clearly demonstrate the added value of using multi-class propagated labels, as \textit{Attention U-Net LP + Man} increases the number of topologically correct segmentations over \textit{Attention U-Net Man} by $68\%$.

\textbf{The top performing experiment regarding the topological correctness of the anomaly area is \textit{Attention U-Net LP + Man + Classifier} (\textbf{DenseMulti})}, our final framework. We find important improvements by adding a classifier to our multi-class framework, particularly regarding cases with a score of 3. These improvements are predominantly DAA cases which initially present a split or unsegmented double arch, corrected by adding a classifier (see  Fig.~\ref{fig:qualitative_results} (right) and Sec.~\ref{subsec:classifier_improv}).

Overall, we still observe some cases with a score of 2, and one with a score of 3 (which includes misclassified subjects) in our full framework (\textit{Attention U-Net LP + Man + Classifier}). We inspect these cases and find that they are the same three subjects with poor segmentation performance across the three experiment training rounds, presenting both poor image quality and misalignments to the atlases. See Appendix~\ref{appendix:qualitative_extended} for extended analysis and visualisation of these cases. 

\begin{figure}[h]
    \centering
    \includegraphics[width=0.65\textwidth]{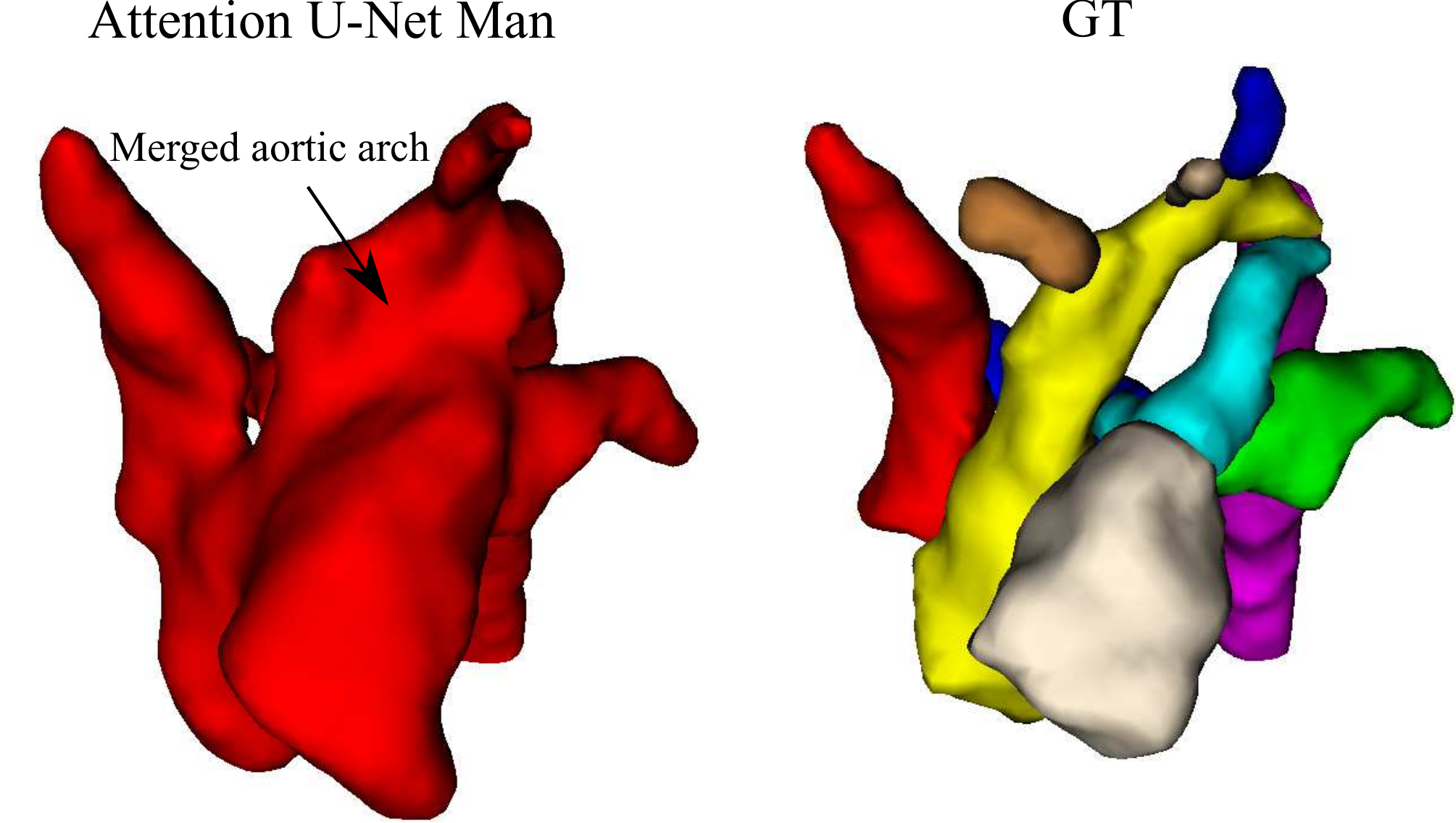}
    \caption{Extreme test set case with aortic merging issue observed.}
    \label{fig:ao_merged_qualitative}
\end{figure}

An important disadvantage of the binary segmentation network is vessel merging, in extreme cases producing an indiscernible aorta (Fig.~\ref{fig:ao_merged_qualitative}). Interestingly, we find the addition of a classifier (\textbf{DenseBin}) to our binary framework (\textit{Attention U-Net Man + Classifier}) to generally degrade performance by heightening this aortic merging problem. Nonetheless, we do observe improvement in certain correctly classed DAA cases (see Sec.~\ref{subsec:classifier_improv}), albeit not being as consistent and robust as our multi-class improvement. 

\section{Discussion}

We present a multi-task deep learning framework for multi-label fetal cardiac vessel segmentation and anomaly classification in T2w 3D fetal MRI.

Fetal CMR has only recently demonstrated its potential \citep{lloyd2019three}, due to the highly specific challenges this data presents. Given this, the field standards regarding automated segmentation performance have yet to be set.  Notably, our dataset requires fast acquisition protocols and motion correction and reconstruction algorithms \citep{uus2020deformable}, due to the inherent motion corruption present in fetal imaging (fetal motion, rapid fetal heartbeat, maternal motion), as well as poor resolution and contrast.

Additionally, there is a range of anatomical variability within each anomaly, as well as 
differing vessel sizes, from the descending aorta which is easily discernible to small head and neck vessels which are sometimes absent even in the GT due to low visibility.

Our specific dataset presents a range of image qualities. We manually score our test set images (based on vessel visibility, artefacts and noise), to provide further explainability into potential failed cases. Our test set comprises 12 high quality subjects, 16 average quality, and 12 low quality images. Therefore, a high segmentation scoring performance is extremely challenging, particularly for small vessels in low quality images, or for cases with important anatomical variability.

\subsection{Segmentation training label type ablation}

Our labelling experiments highlight two important points: (1) adding manual labels substantially increases network performance, and (2) multi-class information is crucial for small vessel detection, as well as anomaly area segmentation. We find significantly increased similarity metrics after including the binary manual labels, with a similar qualitative topology score.  

We find important anatomical inconsistencies in our binary segmentations, including merged vessels, contrary to our proposed multi-class networks. We postulate that this is due to our multi-class network learning each label individually, therefore ensuring that each vessel is present and distinguishable in the final network. The consistency in our propagated atlas labels may also contribute towards this, which contrasts with the inter-observer variability present in the binary manual labels.

The oversegmentation we observed both qualitatively and quantitatively from our binary network is detrimental to clinical usability, as this causes vessel merging which creates indistinguishable vessel topology.

An important point from our analysis is that we find commonly used quantitative metrics to not be fully descriptive of segmentation performance regarding topological and anatomical correctness, hence our qualitative evaluation. Therefore there is a need for future work to explore more fitting topological metrics \citep{hu2019topology}. The inclusion of topology-aware losses \citep{clough2020topological,byrne2021persistent,clough2019explicit,hu2019topology} is also an important aspect for future work to interrogate, given the importance of generating topologically correct predictions for our task. 

Noteworthy, the segmentation task in the present paper could have been tackled following alternative training strategies. A potentially simple solution could be to leverage propagated labels to train a splitting network to convert binary labels to multiple labels. Appendix~\ref{appendix:splitting_networks} contains two additional experiments, that utilise such splitting network. In the first experiment, the splitting network is applied to binary labels predicted by \textit{Attention U-Net Man}, while in the second experiment, the splitting network converts manual binary labels to multi-label segmentations to provide training data for a simple supervised Attention U-Net. We found these strategies to generally lead to more topological errors in vessel segmentations than for our proposed framework, see Fig.~\ref{fig:qualitative_results_splitting}. In particular, we observed more mislabelled vessels in regions with poor visibility due to imaging artefacts.

\subsection{Weak supervision}

The weak supervision ablation study demonstrates that, although our manually generated labels are not fully labelled (as the vessels are represented as a single entity, as opposed to multi-class labels), their addition to the segmenter training framework improves segmentation performance. Even the addition of manually generated labels to only 5\% of the training data (for each diagnosis) supposes a statistically significant improvement for the aortic arch label.

The highest performance is achieved via our proposed segmentation framework: \textit{Attention U-Net LP + Man}. 

\subsection{Classification of anomalies}

We find one explainable mislabelled test case per training round for our \textbf{DenseMulti} classifier (multi-class), and one misclassification overall for \textbf{DenseImgMulti}, with notable improvements in the anomaly area segmentation after joint network training. The addition of a classifier to our multi-class segmentation pipeline consistently improves the topological correctness of all correctly classified DAA cases with segmentation issues, as reflected by our qualitative analysis. 
 It is important to note that the stability of our multi-task training framework depends on accurate hyperparameter tuning of the loss weight balancing in each experiment.

Prior knowledge of the anomaly will always exist in a clinical situation, and therefore misclassifications in \textbf{DenseMulti} aid the identification of cases with faulty segmentation predictions. \textbf{DenseImgMulti}, although being the most accurate classifier, would therefore be less relevant to this application, given its inability to detect topologically incorrect segmentation predictions. It is therefore more suitable as a diagnostic support tool.

An inherent limitation of our approach is the fact that the classifier learns distinctive anomaly-specific features, which restricts the flexibility of the segmentation topology. Therefore, this framework may not generalise well to data which deviate strongly from the training data regarding image quality or anatomical variations. The generalisability of our framework to out-of-distribution cases is therefore a valuable avenue for future work to explore.

The worst performing cases in our test set comprise lower quality SVR reconstructions, that only correct rigid motion, and therefore do not account for deformations in fetal movement, which may result in blurring of the reconstructed 3D images. This was combined with noticable anatomical deviation from the atlases. We find one extreme case where the addition of a classifier accentuated an initial segmentation issue, by fully segmenting the left arch in an RAA case (Fig.~\ref{fig:da_seg_class}). This is an extreme case where we observe the limitation of somewhat topologically constraining our segmentation to a set of learnt distinct features: if there is a low quality case, with poor visibility, the addition of the classifier will push the segmentation towards pertaining to one specific anomaly. Recent advancements in reconstruction techniques \citep{uus2020deformable} may ameliorate this problem, as we find that all of our high quality DSVR reconstructions achieve high accuracy in both segmentation and anomaly classification.

Highly consistent multi-class labels are a key contributor towards our success. We observe a performance degradation in our binary framework  (\textit{Attention U-Net Man + Classifier}), likely due to the more challenging anomaly classifier task: the aorta is not classed as an individual vessel, and the anomaly area occupies a relatively small region in the whole vessels' ROI label. Additionally, the consistency divergence between manually segmented binary labels (inter-observer variability, shortened vessel length for lower quality images) and propagated atlas labels, as well as the lower-quality segmentation predictions of \textit{Attention U-Net Man} (vessel merging and undetected head and neck vessels) contribute towards making our classifier (\textbf{DenseBin}) training data more varied. This impedes the classifier from learning representative aortic arch anomaly features.

Although our work is developed to suit a very specific dataset situation (DSVR and SVR fetal CMR reconstructions, anomaly-specific atlases, and partially labelled training set), our joint classification and segmentation framework may be applicable to many environments with a dual task, or for segmentation improvements in a network trained on distinctly varied data. We note also that our task has highly specific challenges, and therefore success in our application is promising for generalisation to less challenging tasks (e.g. fewer label classes, larger segmentation area, less diverging anomalies). Additionally, our work addresses a partially labelled dataset, largely common in the medical imaging field.

\section{Conclusion}

We present a multi-task approach for joint fetal cardiac vessel segmentation and aortic arch anomaly classification from T2w 3D MRI. We combine deep learning label propagation from anomaly-specific atlases with Attention U-Net segmentation and DenseNet121 classification. We demonstrate a potential application of our segmentation tool for automated diagnosis, by jointly training an anomaly classifier on our output segmentations. Our multi-task approach improves the anomaly area segmentation performance, providing both multi-class segmentations and aortic arch anomaly classification. Our strategy is simple and innovative, and our work has strong clinical applicability, being highly novel regarding our dataset. Our clinical contribution is both to aid vessel visualisation for clinical reporting purposes and to aid diagnostic confidence and efficiency. Our automated multi-task tool may also be useful in training non-expert clinicians, which is challenging due to the highly specific cardiac anomalies, low dataset visibility, and anatomical variability between cases.


\acks{We would like to acknowledge funding from the EPSRC Centre for Doctoral Training in Smart Medical Imaging (EP/S022104/1).

We thank everyone who was involved in the acquisition and examination of the datasets and all participating mothers. This work was supported by the Rosetrees Trust [A2725], the Wellcome/EPSRC Centre for Medical Engineering at King’s College London [WT 203148/Z/16/Z], the Wellcome Trust and EPSRC IEH award [102431] for the iFIND project, the NIHR Clinical Research Facility (CRF) at Guy’s and St Thomas’ and by the National Institute for Health Research Biomedical Research Centre based at Guy’s and St Thomas’ NHS Foundation Trust and King’s College London. The views expressed are those of the authors and not necessarily those of the NHS, the NIHR or the Department of Health.}

%
\ethics{All fetal MRI datasets used in this work were processed subject to informed consent of the participants [REC: 07/H0707/105; REC: 14/LO/1806].

The work follows appropriate ethical standards in conducting research and writing the manuscript, following all applicable laws and regulations regarding treatment of animals or human subjects.}

\coi{We declare we don't have conflicts of interest.}

\data{
The individual fetal MRI datasets used for this study are not publicly available due to ethics regulations.

}

\bibliography{sample}

\begin{thebibliography}{56}
\providecommand{\natexlab}[1]{#1}
\providecommand{\url}[1]{\texttt{#1}}
\expandafter\ifx\csname urlstyle\endcsname\relax
  \providecommand{\doi}[1]{doi: #1}\else
  \providecommand{\doi}{doi: \begingroup \urlstyle{rm}\Url}\fi

\bibitem[Aljabar et~al.(2009)Aljabar, Heckemann, Hammers, Hajnal, and Rueckert]{aljabar2009multi}
Paul Aljabar, Rolf~A Heckemann, Alexander Hammers, Joseph~V Hajnal, and Daniel Rueckert.
\newblock Multi-atlas based segmentation of brain images: atlas selection and its effect on accuracy.
\newblock \emph{Neuroimage}, 46\penalty0 (3):\penalty0 726--738, 2009.

\bibitem[Arafati et~al.(2019)Arafati, Hu, Finn, Rickers, Cheng, Jafarkhani, and Kheradvar]{arafati2019artificial}
Arghavan Arafati, Peng Hu, J~Paul Finn, Carsten Rickers, Andrew~L Cheng, Hamid Jafarkhani, and Arash Kheradvar.
\newblock Artificial intelligence in pediatric and adult congenital cardiac mri: an unmet clinical need.
\newblock \emph{Cardiovascular diagnosis and therapy}, 9\penalty0 (Suppl 2):\penalty0 S310, 2019.

\bibitem[Balakrishnan et~al.(2019)Balakrishnan, Zhao, Sabuncu, Guttag, and Dalca]{balakrishnan2019voxelmorph}
Guha Balakrishnan, Amy Zhao, Mert~R Sabuncu, John Guttag, and Adrian~V Dalca.
\newblock Voxelmorph: a learning framework for deformable medical image registration.
\newblock \emph{IEEE transactions on medical imaging}, 38\penalty0 (8):\penalty0 1788--1800, 2019.

\bibitem[Brown et~al.(2006)Brown, Ridout, Hoskote, Verhulst, Ricci, and Bull]{brown2006delayed}
Kate~L Brown, Deborah~A Ridout, Aparna Hoskote, Lynda Verhulst, Marco Ricci, and Catherine Bull.
\newblock Delayed diagnosis of congenital heart disease worsens preoperative condition and outcome of surgery in neonates.
\newblock \emph{Heart}, 92\penalty0 (9):\penalty0 1298--1302, 2006.

\bibitem[Byrne et~al.(2021)Byrne, Clough, Montana, and King]{byrne2021persistent}
Nick Byrne, James~R Clough, Giovanni Montana, and Andrew~P King.
\newblock A persistent homology-based topological loss function for multi-class cnn segmentation of cardiac mri.
\newblock In \emph{Statistical Atlases and Computational Models of the Heart. M\&Ms and EMIDEC Challenges: 11th International Workshop, STACOM 2020, Held in Conjunction with MICCAI 2020, Lima, Peru, October 4, 2020, Revised Selected Papers 11}, pages 3--13. Springer, 2021.

\bibitem[Chen et~al.(2020)Chen, Qin, Qiu, Tarroni, Duan, Bai, and Rueckert]{chen2020deep}
Chen Chen, Chen Qin, Huaqi Qiu, Giacomo Tarroni, Jinming Duan, Wenjia Bai, and Daniel Rueckert.
\newblock Deep learning for cardiac image segmentation: a review.
\newblock \emph{Frontiers in Cardiovascular Medicine}, 7:\penalty0 25, 2020.

\bibitem[Clough et~al.(2019)Clough, Oksuz, Byrne, Schnabel, and King]{clough2019explicit}
James~R Clough, Ilkay Oksuz, Nicholas Byrne, Julia~A Schnabel, and Andrew~P King.
\newblock Explicit topological priors for deep-learning based image segmentation using persistent homology.
\newblock In \emph{Information Processing in Medical Imaging: 26th International Conference, IPMI 2019, Hong Kong, China, June 2--7, 2019, Proceedings 26}, pages 16--28. Springer, 2019.

\bibitem[Clough et~al.(2020)Clough, Byrne, Oksuz, Zimmer, Schnabel, and King]{clough2020topological}
James~R Clough, Nicholas Byrne, Ilkay Oksuz, Veronika~A Zimmer, Julia~A Schnabel, and Andrew~P King.
\newblock A topological loss function for deep-learning based image segmentation using persistent homology.
\newblock \emph{IEEE Transactions on Pattern Analysis and Machine Intelligence}, 44\penalty0 (12):\penalty0 8766--8778, 2020.

\bibitem[DeVore et~al.(2003)DeVore, Falkensammer, Sklansky, and Platt]{devore2003spatio}
GR~DeVore, P~Falkensammer, MS~Sklansky, and LD~Platt.
\newblock Spatio-temporal image correlation (stic): new technology for evaluation of the fetal heart.
\newblock \emph{Ultrasound in Obstetrics and Gynecology: The Official Journal of the International Society of Ultrasound in Obstetrics and Gynecology}, 22\penalty0 (4):\penalty0 380--387, 2003.

\bibitem[Dinsdale et~al.(2019)Dinsdale, Jenkinson, and Namburete]{dinsdale2019spatial}
Nicola~K Dinsdale, Mark Jenkinson, and Ana~IL Namburete.
\newblock Spatial warping network for {3D} segmentation of the hippocampus in mr images.
\newblock In \emph{International Conference on Medical Image Computing and Computer-Assisted Intervention}, pages 284--291. Springer, 2019.

\bibitem[Dong and Zhu(2018)]{dong2018utility}
Su-Zhen Dong and Ming Zhu.
\newblock Utility of fetal cardiac magnetic resonance imaging to assess fetuses with right aortic arch and right ductus arteriosus.
\newblock \emph{The Journal of Maternal-Fetal \& Neonatal Medicine}, 31\penalty0 (12):\penalty0 1627--1631, 2018.

\bibitem[Dong et~al.(2020)Dong, Zhu, Ji, Ren, and Liu]{dong2020fetal}
Su-Zhen Dong, Ming Zhu, Hui Ji, Jing-Ya Ren, and Ke~Liu.
\newblock Fetal cardiac mri: a single center experience over 14-years on the potential utility as an adjunct to fetal technically inadequate echocardiography.
\newblock \emph{Scientific Reports}, 10\penalty0 (1):\penalty0 1--10, 2020.

\bibitem[Ebner et~al.(2020)Ebner, Wang, Li, Aertsen, Patel, Aughwane, Melbourne, Doel, Dymarkowski, De~Coppi, et~al.]{ebner2020automated}
Michael Ebner, Guotai Wang, Wenqi Li, Michael Aertsen, Premal~A Patel, Rosalind Aughwane, Andrew Melbourne, Tom Doel, Steven Dymarkowski, Paolo De~Coppi, et~al.
\newblock An automated framework for localization, segmentation and super-resolution reconstruction of fetal brain {MRI}.
\newblock \emph{NeuroImage}, 206:\penalty0 116324, 2020.

\bibitem[Grigorescu et~al.(2020)Grigorescu, Uus, Christiaens, Cordero-Grande, Hutter, Edwards, Hajnal, Modat, and Deprez]{grigorescu2020diffusion}
Irina Grigorescu, Alena Uus, Daan Christiaens, Lucilio Cordero-Grande, Jana Hutter, A~David Edwards, Joseph~V Hajnal, Marc Modat, and Maria Deprez.
\newblock Diffusion tensor driven image registration: a deep learning approach.
\newblock In \emph{International Workshop on Biomedical Image Registration}, pages 131--140. Springer, 2020.

\bibitem[Hatamizadeh et~al.(2022)Hatamizadeh, Tang, Nath, Yang, Myronenko, Landman, Roth, and Xu]{hatamizadeh2022unetr}
Ali Hatamizadeh, Yucheng Tang, Vishwesh Nath, Dong Yang, Andriy Myronenko, Bennett Landman, Holger~R Roth, and Daguang Xu.
\newblock Unetr: Transformers for {3D} medical image segmentation.
\newblock In \emph{Proceedings of the IEEE/CVF Winter Conference on Applications of Computer Vision}, pages 574--584, 2022.

\bibitem[Heckemann et~al.(2006)Heckemann, Hajnal, Aljabar, Rueckert, and Hammers]{heckemann2006automatic}
Rolf~A Heckemann, Joseph~V Hajnal, Paul Aljabar, Daniel Rueckert, and Alexander Hammers.
\newblock Automatic anatomical brain mri segmentation combining label propagation and decision fusion.
\newblock \emph{NeuroImage}, 33\penalty0 (1):\penalty0 115--126, 2006.

\bibitem[Hesamian et~al.(2019)Hesamian, Jia, He, and Kennedy]{hesamian2019deep}
Mohammad~Hesam Hesamian, Wenjing Jia, Xiangjian He, and Paul Kennedy.
\newblock Deep learning techniques for medical image segmentation: achievements and challenges.
\newblock \emph{Journal of digital imaging}, 32\penalty0 (4):\penalty0 582--596, 2019.

\bibitem[Hu et~al.(2019)Hu, Li, Samaras, and Chen]{hu2019topology}
Xiaoling Hu, Fuxin Li, Dimitris Samaras, and Chao Chen.
\newblock Topology-preserving deep image segmentation.
\newblock \emph{Advances in neural information processing systems}, 32, 2019.

\bibitem[Huang et~al.(2017)Huang, Liu, Van Der~Maaten, and Weinberger]{huang2017densely}
Gao Huang, Zhuang Liu, Laurens Van Der~Maaten, and Kilian~Q Weinberger.
\newblock Densely connected convolutional networks.
\newblock In \emph{Proceedings of the IEEE conference on computer vision and pattern recognition}, pages 4700--4708, 2017.

\bibitem[Isensee et~al.(2018)Isensee, Petersen, Klein, Zimmerer, Jaeger, Kohl, Wasserthal, Koehler, Norajitra, Wirkert, et~al.]{isensee2018nnu}
Fabian Isensee, Jens Petersen, Andre Klein, David Zimmerer, Paul~F Jaeger, Simon Kohl, Jakob Wasserthal, Gregor Koehler, Tobias Norajitra, Sebastian Wirkert, et~al.
\newblock nnu-net: Self-adapting framework for u-net-based medical image segmentation.
\newblock \emph{arXiv preprint arXiv:1809.10486}, 2018.

\bibitem[Kainz et~al.(2015)Kainz, Steinberger, Wein, Kuklisova-Murgasova, Malamateniou, Keraudren, Torsney-Weir, Rutherford, Aljabar, Hajnal, et~al.]{kainz2015fast}
Bernhard Kainz, Markus Steinberger, Wolfgang Wein, Maria Kuklisova-Murgasova, Christina Malamateniou, Kevin Keraudren, Thomas Torsney-Weir, Mary Rutherford, Paul Aljabar, Joseph~V Hajnal, et~al.
\newblock Fast volume reconstruction from motion corrupted stacks of 2d slices.
\newblock \emph{IEEE transactions on medical imaging}, 34\penalty0 (9):\penalty0 1901--1913, 2015.

\bibitem[Keraudren et~al.(2014)Keraudren, Kuklisova-Murgasova, Kyriakopoulou, Malamateniou, Rutherford, Kainz, Hajnal, and Rueckert]{keraudren2014automated}
Kevin Keraudren, Maria Kuklisova-Murgasova, Vanessa Kyriakopoulou, Christina Malamateniou, Mary~A Rutherford, Bernhard Kainz, Joseph~V Hajnal, and Daniel Rueckert.
\newblock Automated fetal brain segmentation from 2d {MRI} slices for motion correction.
\newblock \emph{NeuroImage}, 101:\penalty0 633--643, 2014.

\bibitem[Khalili et~al.(2019)Khalili, Lessmann, Turk, Claessens, de~Heus, Kolk, Viergever, Benders, and I{\v{s}}gum]{khalili2019automatic}
Nadieh Khalili, Nikolas Lessmann, Elise Turk, N~Claessens, Roel de~Heus, Tessel Kolk, Max~A Viergever, Manon~JNL Benders, and Ivana I{\v{s}}gum.
\newblock Automatic brain tissue segmentation in fetal {MRI} using convolutional neural networks.
\newblock \emph{Magnetic resonance imaging}, 64:\penalty0 77--89, 2019.

\bibitem[Kuklisova-Murgasova et~al.(2012)Kuklisova-Murgasova, Quaghebeur, Rutherford, Hajnal, and Schnabel]{kuklisova2012reconstruction}
Maria Kuklisova-Murgasova, Gerardine Quaghebeur, Mary~A Rutherford, Joseph~V Hajnal, and Julia~A Schnabel.
\newblock Reconstruction of fetal brain {MRI} with intensity matching and complete outlier removal.
\newblock \emph{Medical image analysis}, 16\penalty0 (8):\penalty0 1550--1564, 2012.

\bibitem[Liao et~al.(2016)Liao, Kodagoda, Wang, Shi, and Liu]{liao2016understand}
Yiyi Liao, Sarath Kodagoda, Yue Wang, Lei Shi, and Yong Liu.
\newblock Understand scene categories by objects: A semantic regularized scene classifier using convolutional neural networks.
\newblock In \emph{2016 IEEE international conference on robotics and automation (ICRA)}, pages 2318--2325. IEEE, 2016.

\bibitem[Lin et~al.(2017)Lin, Goyal, Girshick, He, and Doll{\'a}r]{lin2017focal}
Tsung-Yi Lin, Priya Goyal, Ross Girshick, Kaiming He, and Piotr Doll{\'a}r.
\newblock Focal loss for dense object detection.
\newblock In \emph{Proceedings of the IEEE international conference on computer vision}, pages 2980--2988, 2017.

\bibitem[Lloyd et~al.(2019)Lloyd, Pushparajah, Simpson, Van~Amerom, Van~Poppel, Schulz, Kainz, Deprez, Lohezic, Allsop, et~al.]{lloyd2019three}
David~FA Lloyd, Kuberan Pushparajah, John~M Simpson, Joshua~FP Van~Amerom, Milou~PM Van~Poppel, Alexander Schulz, Bernard Kainz, Maria Deprez, Maelene Lohezic, Joanna Allsop, et~al.
\newblock Three-dimensional visualisation of the fetal heart using prenatal {MRI} with motion-corrected slice-volume registration: a prospective, single-centre cohort study.
\newblock \emph{The Lancet}, 393\penalty0 (10181):\penalty0 1619--1627, 2019.

\bibitem[Mazwi et~al.(2013)Mazwi, Brown, Marshall, Pigula, Laussen, Polito, Wypij, and Costello]{mazwi2013unplanned}
Mjaye~L Mazwi, David~W Brown, Audrey~C Marshall, Frank~A Pigula, Peter~C Laussen, Angelo Polito, David Wypij, and John~M Costello.
\newblock Unplanned reinterventions are associated with postoperative mortality in neonates with critical congenital heart disease.
\newblock \emph{The Journal of thoracic and cardiovascular surgery}, 145\penalty0 (3):\penalty0 671--677, 2013.

\bibitem[Mehta et~al.(2018)Mehta, Mercan, Bartlett, Weaver, Elmore, and Shapiro]{mehta2018net}
Sachin Mehta, Ezgi Mercan, Jamen Bartlett, Donald Weaver, Joann~G Elmore, and Linda Shapiro.
\newblock Y-net: joint segmentation and classification for diagnosis of breast biopsy images.
\newblock In \emph{International Conference on Medical Image Computing and Computer-Assisted Intervention}, pages 893--901. Springer, 2018.

\bibitem[Mendis et~al.(2011)Mendis, Puska, Norrving, Organization, et~al.]{mendis2011global}
Shanthi Mendis, Pekka Puska, Bo~Norrving, World~Health Organization, et~al.
\newblock \emph{Global atlas on cardiovascular disease prevention and control}.
\newblock World Health Organization, 2011.

\bibitem[Oktay et~al.(2018)Oktay, Schlemper, Folgoc, Lee, Heinrich, Misawa, Mori, McDonagh, Hammerla, Kainz, et~al.]{oktay2018attention}
Ozan Oktay, Jo~Schlemper, Loic~Le Folgoc, Matthew Lee, Mattias Heinrich, Kazunari Misawa, Kensaku Mori, Steven McDonagh, Nils~Y Hammerla, Bernhard Kainz, et~al.
\newblock Attention u-net: Learning where to look for the pancreas.
\newblock \emph{arXiv preprint arXiv:1804.03999}, 2018.

\bibitem[Patel et~al.(1997)Patel, Klufas, Alberico, and Edelman]{patel1997half}
Mahesh~R Patel, Roman~A Klufas, Ronald~A Alberico, and Robert~R Edelman.
\newblock Half-fourier acquisition single-shot turbo spin-echo (haste) mr: comparison with fast spin-echo mr in diseases of the brain.
\newblock \emph{American journal of neuroradiology}, 18\penalty0 (9):\penalty0 1635--1640, 1997.

\bibitem[Payette et~al.(2020)Payette, Kottke, and Jakab]{payette2020efficient}
Kelly Payette, Raimund Kottke, and Andras Jakab.
\newblock Efficient multi-class fetal brain segmentation in high resolution {MRI} reconstructions with noisy labels.
\newblock In \emph{Medical Ultrasound, and Preterm, Perinatal and Paediatric Image Analysis}, pages 295--304. Springer, 2020.

\bibitem[Peng and Wang(2021)]{peng2021medical}
Jialin Peng and Ye~Wang.
\newblock Medical image segmentation with limited supervision: a review of deep network models.
\newblock \emph{IEEE Access}, 9:\penalty0 36827--36851, 2021.

\bibitem[Puyol-Ant{\'o}n et~al.(2021)Puyol-Ant{\'o}n, Ruijsink, Piechnik, Neubauer, Petersen, Razavi, and King]{puyol2021fairness}
Esther Puyol-Ant{\'o}n, Bram Ruijsink, Stefan~K Piechnik, Stefan Neubauer, Steffen~E Petersen, Reza Razavi, and Andrew~P King.
\newblock Fairness in cardiac mr image analysis: an investigation of bias due to data imbalance in deep learning based segmentation.
\newblock In \emph{International Conference on Medical Image Computing and Computer-Assisted Intervention}, pages 413--423. Springer, 2021.

\bibitem[Ramirez~Gilliland et~al.(2022)Ramirez~Gilliland, Uus, van Poppel, Grigorescu, Steinweg, Lloyd, Pushparajah, King, and Deprez]{ramirez2022automated}
Paula Ramirez~Gilliland, Alena Uus, Milou~PM van Poppel, Irina Grigorescu, Johannes~K Steinweg, David~FA Lloyd, Kuberan Pushparajah, Andrew~P King, and Maria Deprez.
\newblock Automated multi-class fetal cardiac vessel segmentation in aortic arch anomalies using t2-weighted 3d fetal mri.
\newblock In \emph{International Workshop on Preterm, Perinatal and Paediatric Image Analysis}, pages 82--93. Springer, 2022.

\bibitem[Rezaei et~al.(2018)Rezaei, Yang, and Meinel]{rezaei2018whole}
Mina Rezaei, Haojin Yang, and Christoph Meinel.
\newblock Whole heart and great vessel segmentation with context-aware of generative adversarial networks.
\newblock In \emph{Bildverarbeitung f{\"u}r die Medizin 2018}, pages 353--358. Springer, 2018.

\bibitem[Ronneberger et~al.(2015)Ronneberger, Fischer, and Brox]{ronneberger2015u}
Olaf Ronneberger, Philipp Fischer, and Thomas Brox.
\newblock U-net: Convolutional networks for biomedical image segmentation.
\newblock In \emph{International Conference on Medical image computing and computer-assisted intervention}, pages 234--241. Springer, 2015.

\bibitem[Rueckert et~al.(1999{\natexlab{a}})Rueckert, Sonoda, Denton, Rankin, Hayes, Leach, Hill, and Hawkes]{rueckert1999comparison}
Daniel Rueckert, Luke~I Sonoda, Erica~RE Denton, S~Rankin, Carmel Hayes, Martin~O Leach, Derek~LG Hill, and David~John Hawkes.
\newblock Comparison and evaluation of rigid and nonrigid registration of breast mr images.
\newblock In \emph{Medical Imaging 1999: Image Processing}, volume 3661, pages 78--88. International Society for Optics and Photonics, 1999{\natexlab{a}}.

\bibitem[Rueckert et~al.(1999{\natexlab{b}})Rueckert, Sonoda, Hayes, Hill, Leach, and Hawkes]{rueckert1999nonrigid}
Daniel Rueckert, Luke~I Sonoda, Carmel Hayes, Derek~LG Hill, Martin~O Leach, and David~J Hawkes.
\newblock Nonrigid registration using free-form deformations: application to breast mr images.
\newblock \emph{IEEE transactions on medical imaging}, 18\penalty0 (8):\penalty0 712--721, 1999{\natexlab{b}}.

\bibitem[Salehi et~al.(2021)Salehi, Fricke, Bhat, Arheden, Liuba, and Hedstr{\"o}m]{salehi2021utility}
Daniel Salehi, Katrin Fricke, Misha Bhat, H{\aa}kan Arheden, Petru Liuba, and Erik Hedstr{\"o}m.
\newblock Utility of fetal cardiovascular magnetic resonance for prenatal diagnosis of complex congenital heart defects.
\newblock \emph{JAMA network open}, 4\penalty0 (3):\penalty0 e213538--e213538, 2021.

\bibitem[Salehi et~al.(2018)Salehi, Hashemi, Velasco-Annis, Ouaalam, Estroff, Erdogmus, Warfield, and Gholipour]{salehi2018real}
Seyed Sadegh~Mohseni Salehi, Seyed~Raein Hashemi, Clemente Velasco-Annis, Abdelhakim Ouaalam, Judy~A Estroff, Deniz Erdogmus, Simon~K Warfield, and Ali Gholipour.
\newblock Real-time automatic fetal brain extraction in fetal {MRI} by deep learning.
\newblock In \emph{2018 IEEE 15th International Symposium on Biomedical Imaging (ISBI 2018)}, pages 720--724. IEEE, 2018.

\bibitem[Schnabel et~al.(2001)Schnabel, Rueckert, Quist, Blackall, Castellano-Smith, Hartkens, Penney, Hall, Liu, Truwit, et~al.]{schnabel2001generic}
Julia~A Schnabel, Daniel Rueckert, Marcel Quist, Jane~M Blackall, Andy~D Castellano-Smith, Thomas Hartkens, Graeme~P Penney, Walter~A Hall, Haiying Liu, Charles~L Truwit, et~al.
\newblock A generic framework for non-rigid registration based on non-uniform multi-level free-form deformations.
\newblock In \emph{International Conference on Medical Image Computing and Computer-Assisted Intervention}, pages 573--581. Springer, 2001.

\bibitem[Semelka et~al.(1996)Semelka, Kelekis, Thomasson, Brown, and Laub]{semelka1996haste}
Richard~C Semelka, Nikolaos~L Kelekis, David Thomasson, Mark~A Brown, and Gerhard~A Laub.
\newblock Haste mr imaging: description of technique and preliminary results in the abdomen.
\newblock \emph{Journal of Magnetic Resonance Imaging}, 6\penalty0 (4):\penalty0 698--699, 1996.

\bibitem[Sinclair et~al.(2022)Sinclair, Schuh, Hahn, Petersen, Bai, Batten, Schaap, and Glocker]{sinclair2022atlas}
Matthew Sinclair, Andreas Schuh, Karl Hahn, Kersten Petersen, Ying Bai, James Batten, Michiel Schaap, and Ben Glocker.
\newblock Atlas-istn: Joint segmentation, registration and atlas construction with image-and-spatial transformer networks.
\newblock \emph{Medical Image Analysis}, 78:\penalty0 102383, 2022.

\bibitem[Sudre et~al.(2017)Sudre, Li, Vercauteren, Ourselin, and Cardoso]{sudre2017generalised}
Carole~H Sudre, Wenqi Li, Tom Vercauteren, Sebastien Ourselin, and M~Jorge Cardoso.
\newblock Generalised dice overlap as a deep learning loss function for highly unbalanced segmentations.
\newblock In \emph{Deep learning in medical image analysis and multimodal learning for clinical decision support}, pages 240--248. Springer, 2017.

\bibitem[Teichmann et~al.(2018)Teichmann, Weber, Zoellner, Cipolla, and Urtasun]{teichmann2018multinet}
Marvin Teichmann, Michael Weber, Marius Zoellner, Roberto Cipolla, and Raquel Urtasun.
\newblock Multinet: Real-time joint semantic reasoning for autonomous driving.
\newblock In \emph{2018 IEEE intelligent vehicles symposium (IV)}, pages 1013--1020. IEEE, 2018.

\bibitem[Uus et~al.(2020)Uus, Zhang, Jackson, Roberts, Rutherford, Hajnal, and Deprez]{uus2020deformable}
Alena Uus, Tong Zhang, Laurence~H Jackson, Thomas~A Roberts, Mary~A Rutherford, Joseph~V Hajnal, and Maria Deprez.
\newblock Deformable slice-to-volume registration for motion correction of fetal body and placenta mri.
\newblock \emph{IEEE transactions on medical imaging}, 39\penalty0 (9):\penalty0 2750--2759, 2020.

\bibitem[Uus et~al.(2021)Uus, Grigorescu, van Poppel, Hughes, Steinweg, Roberts, Lloyd, Pushparajah, and Deprez]{uus20213d}
Alena Uus, Irina Grigorescu, Milou van Poppel, Emer Hughes, Johannes Steinweg, Thomas Roberts, David Lloyd, Kuberan Pushparajah, and Maria Deprez.
\newblock {3D} unet with gan discriminator for robust localisation of the fetal brain and trunk in {MRI} with partial coverage of the fetal body.
\newblock \emph{bioRxiv}, 2021.

\bibitem[Uus et~al.(2022{\natexlab{a}})Uus, Grigorescu, {van Poppel}, Steinweg, Roberts, Rutherford, Hajnal, Lloyd, Pushparajah, and Deprez]{uus2022media}
Alena~U. Uus, Irina Grigorescu, Milou~P.M. {van Poppel}, Johannes~K. Steinweg, Thomas~A. Roberts, Mary~A. Rutherford, Joseph~V. Hajnal, David~F.A. Lloyd, Kuberan Pushparajah, and Maria Deprez.
\newblock Automated 3d reconstruction of the fetal thorax in the standard atlas space from motion-corrupted mri stacks for 21–36 weeks ga range.
\newblock \emph{Medical Image Analysis}, 80:\penalty0 102484, 2022{\natexlab{a}}.
\newblock ISSN 1361-8415.
\newblock \doi{https://doi.org/10.1016/j.media.2022.102484}.
\newblock URL \url{https://www.sciencedirect.com/science/article/pii/S1361841522001311}.

\bibitem[Uus et~al.(2022{\natexlab{b}})Uus, van Poppel, Steinweg, Grigorescu, Ramirez~Gilliland, Roberts, Egloff~Collado, Rutherford, Hajnal, Lloyd, et~al.]{uus20223d}
Alena~U Uus, Milou~PM van Poppel, Johannes~K Steinweg, Irina Grigorescu, Paula Ramirez~Gilliland, Thomas~A Roberts, Alexia Egloff~Collado, Mary~A Rutherford, Joseph~V Hajnal, David~FA Lloyd, et~al.
\newblock 3d black blood cardiovascular magnetic resonance atlases of congenital aortic arch anomalies and the normal fetal heart: application to automated multi-label segmentation.
\newblock \emph{Journal of Cardiovascular Magnetic Resonance}, 24\penalty0 (1):\penalty0 1--13, 2022{\natexlab{b}}.

\bibitem[Xu et~al.(2019)Xu, Wang, Shi, Yuan, Jia, Huang, and Zhuang]{xu2019whole}
Xiaowei Xu, Tianchen Wang, Yiyu Shi, Haiyun Yuan, Qianjun Jia, Meiping Huang, and Jian Zhuang.
\newblock Whole heart and great vessel segmentation in congenital heart disease using deep neural networks and graph matching.
\newblock In \emph{International Conference on Medical Image Computing and Computer-Assisted Intervention}, pages 477--485. Springer, 2019.

\bibitem[Xu and Niethammer(2019)]{xu2019deepatlas}
Zhenlin Xu and Marc Niethammer.
\newblock Deepatlas: Joint semi-supervised learning of image registration and segmentation.
\newblock In \emph{International Conference on Medical Image Computing and Computer-Assisted Intervention}, pages 420--429. Springer, 2019.

\bibitem[Yu et~al.(2016)Yu, Yang, Qin, and Heng]{yu20163d}
Lequan Yu, Xin Yang, Jing Qin, and Pheng-Ann Heng.
\newblock {3D} fractalnet: dense volumetric segmentation for cardiovascular {MRI} volumes.
\newblock In \emph{Reconstruction, segmentation, and analysis of medical images}, pages 103--110. Springer, 2016.

\bibitem[Yushkevich et~al.(2006)Yushkevich, Piven, Cody~Hazlett, Gimpel~Smith, Ho, Gee, and Gerig]{py06nimg}
Paul~A. Yushkevich, Joseph Piven, Heather Cody~Hazlett, Rachel Gimpel~Smith, Sean Ho, James~C. Gee, and Guido Gerig.
\newblock User-guided {3D} active contour segmentation of anatomical structures: Significantly improved efficiency and reliability.
\newblock \emph{Neuroimage}, 31\penalty0 (3):\penalty0 1116--1128, 2006.

\bibitem[Zhao et~al.(2019)Zhao, Balakrishnan, Durand, Guttag, and Dalca]{zhao2019data}
Amy Zhao, Guha Balakrishnan, Fredo Durand, John~V Guttag, and Adrian~V Dalca.
\newblock Data augmentation using learned transformations for one-shot medical image segmentation.
\newblock In \emph{Proceedings of the IEEE/CVF conference on computer vision and pattern recognition}, pages 8543--8553, 2019.

\end{thebibliography}


\clearpage
\appendix

\section{Alternative segmentation strategies}
\label{appendix:splitting_networks}

Here we present two additional approaches to our proposed segmentation strategy. These represent the alternative way of combining our dataset labels (multi-class propagated labels and manually generated binary labels). The strategy behind these new approaches is to learn to split binary labels into multi-class labels. We examine the following:

\begin{enumerate}
    \item{\textbf{\textit{Attention U-Net Man + Split}}: Attention U-Net that splits the predicted CNN binary labels from Attention U-Net Man into multi-class labels.}
    \item{\textbf{\textit{Attention U-Net Split}}: two-step strategy, that involves first splitting the manually segmented binary labels into multi-class (with a CNN), followed by training an additional Attention U-Net using these split labels.}
\end{enumerate}

Both these approaches rely on a trained splitting network. This splitting network is trained on joined propagated labels as input (binary) and learns to split these into multi-class, thereby learning to preserve the shape of the input segmentation.

\begin{figure}[h]
    \centering
    \includegraphics[width=0.9\columnwidth]{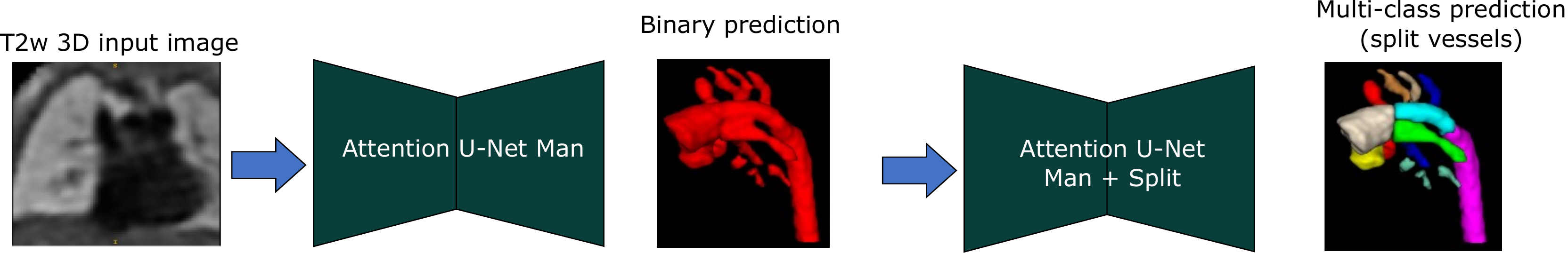}
    \caption{Inference time predictions for \textit{Attention U-Net Man + Split}, which does not require any labels. A binary prediction is generated for the input image, which is then split.}
    \label{fig:diagram_netmansplit}
\end{figure}

For \textit{Attention U-Net Man + Split}, we use \textit{Attention U-Net Man} to generate the binary labels at inference time, see Fig.~\ref{fig:diagram_netmansplit}.

In accordance with our other segmentation experiments, we repeat each experiment three times, to account for network stochasticity.

\subsection{Quantitative Results}
An important training disadvantage of these two approaches is that they require twice the computational time compared to our proposed method (\textit{Attention U-Net LP + Man}), given that two CNNs are required instead of just one.

\begin{figure}[h]
     \centering
     \begin{subfigure}[b]{0.496\textwidth}
         \centering
         \includegraphics[width=\textwidth]{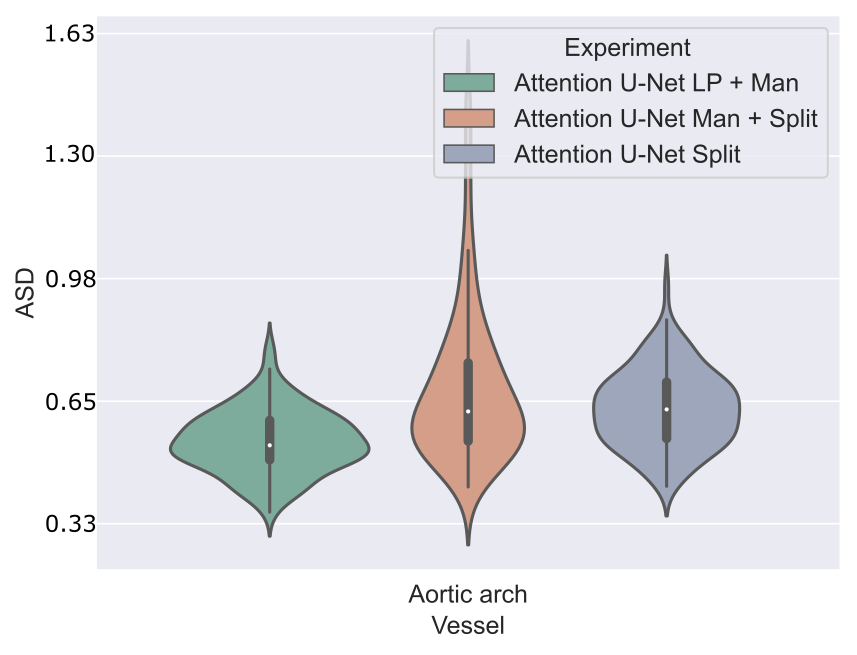}
         \caption{ASD scores (lower = better).}
         \label{fig:asd_split_lpman}
     \end{subfigure}
     \hfill
     \begin{subfigure}[b]{0.496\textwidth}
         \centering
         \includegraphics[width=\textwidth]{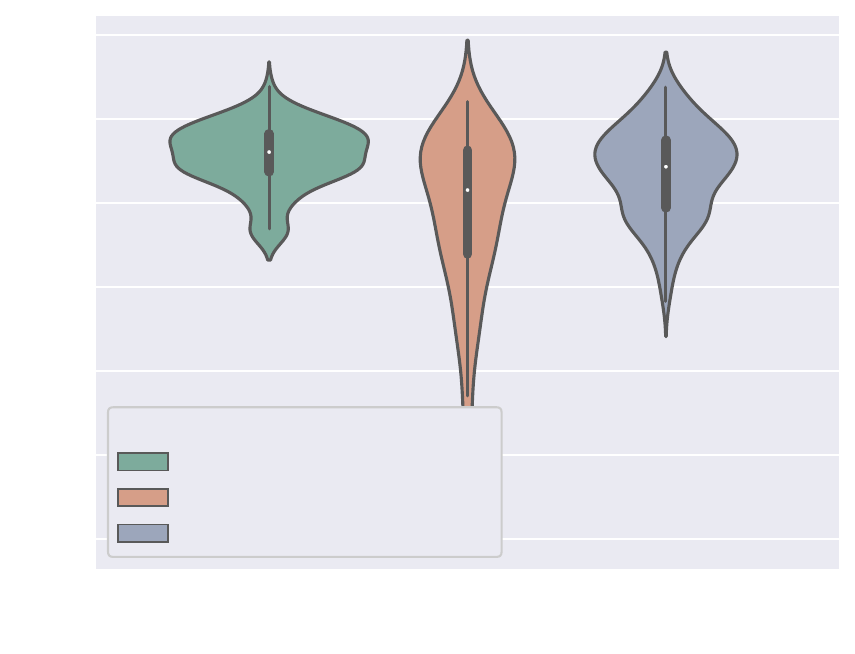}
         \caption{Dice scores (higher = better).}
         \label{fig:dice_split_manlp}
     \end{subfigure}

           \caption{Quantitative scores for the aortic arch (anomaly area) on test set comparing our proposed labelling type training combination (\textit{Attention U-Net LP + Man}) to \textit{Attention U-Net Man + Split}. We repeat each experiment three times and include all results here.}
        \label{fig:split_violin}
\end{figure}

\textit{Attention U-Net Man + LP} and \textit{Attention U-Net Man + Split} are compared in Fig.~\ref{fig:split_violin} for the aortic arch (key anomaly area). These results display more extreme results for \textit{Attention U-Net Man + Split}. A Mann-Whitney U test reveals a statistically significant difference (p-value=$8.23\times 10^{-9}$) between \textit{Attention U-Net LP + Man} and \textit{Attention U-Net Man + Split}, however not between \textit{Attention U-Net Split} and \textit{Attention U-Net LP + Man}.

\subsection{Qualitative Results}

\begin{figure}[h]
    \centering
    \includegraphics[width=\textwidth]{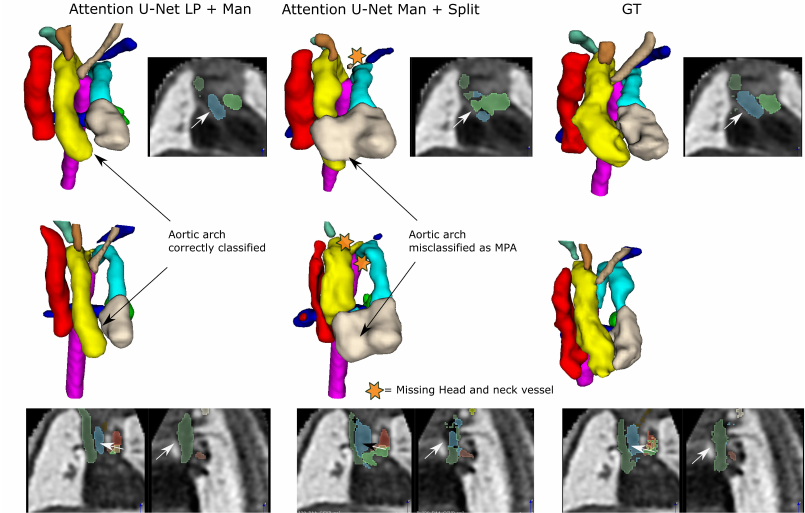}
    \caption{Vessel misclassification issue observed in \textit{Attention U-Net Man + Split}, with comparison to our proposed labelling combination \textit{Attention U-Net LP + Man} and manually segmented GT. Predictions are shown for two independent test set cases. Orange stars indicate missing vessels in the splitting network, and white arrows (T2w images) indicate the misclassification area.}
    \label{fig:split_misclass}
    
\end{figure}

The inherent topological issues observed in the binary network (\textit{Attention U-Net Man}) are propagated to its subsequent segmentation splitting variant, \textit{Attention U-Net Man + Split}. This causes vessel misclassification, as displayed in two independent test set cases in Fig.~\ref{fig:split_misclass}. Undetected small vessels are also signalled.

\textit{Attention U-Net Man + Split} takes the output of \textit{Attention U-Net Man} as input and performs vessel splitting, learning to preserve the shape of the input segmentation. This leads to the perpetuation of the existing topological problems in the predicted segmentation. Moreover, when vessel merging occurs, the splitting network further hampers performance by misclassifying merged vessels. This vessel misclassification issue is not as persistent in \textit{Attention U-Net Split}, the network which is trained on split manual labels, however still occurs on lower quality cases, particularly cases presenting challenging discernment between aorta and MPA. 

\begin{figure}[h]
    \centering
    \includegraphics[width=\textwidth]{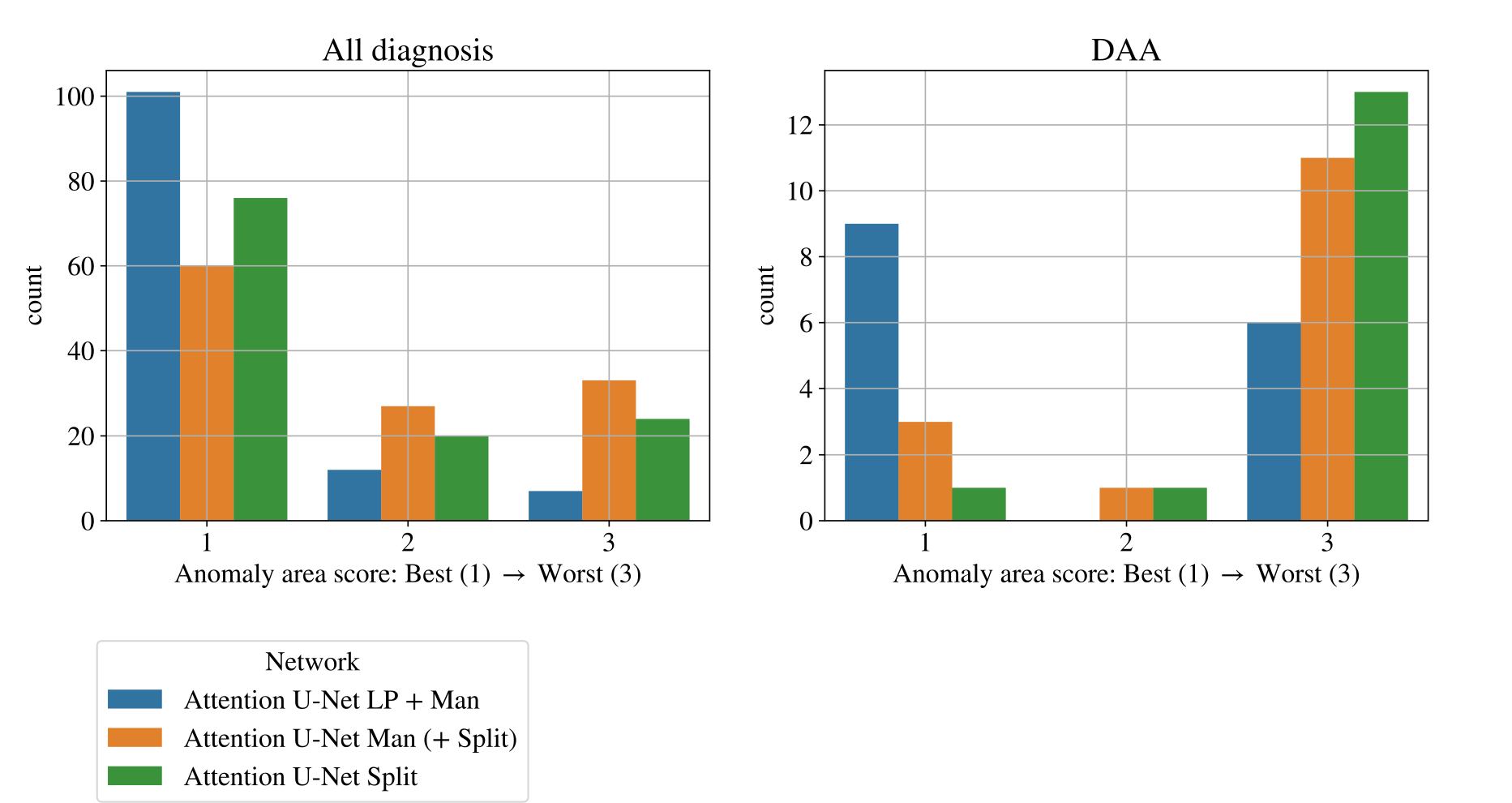}
    \caption{Qualitative analysis results on topology correctness of the anomaly area (aortic arch) on our test set for \textbf{all diagnoses} (left), and for DAA cases only (right). Each training experiment was repeated three times, and we include all of our results here (i.e. we have three results for each of our 40 test set cases, one for each round, so the total number of cases is 120)}. 
    \label{fig:qualitative_results_splitting}
\end{figure}

We quantify this by conducting a qualitative assessment of the segmentation topology area (see Section~\ref{sec:qualitative_analysis_definition} for definition). We include scores for all vessels and for the anomaly area in Fig.~\ref{fig:qualitative_results_splitting}. This showcases the superior performance of our proposed method, followed by \textit{Attention U-Net Split}.

\section{Anomaly classifier}
\label{appendix:condition_classifier}

\begin{figure}[h]
     \centering
     \begin{subfigure}{0.49\textwidth}
         \centering 
         \includegraphics[width=1.1\textwidth]{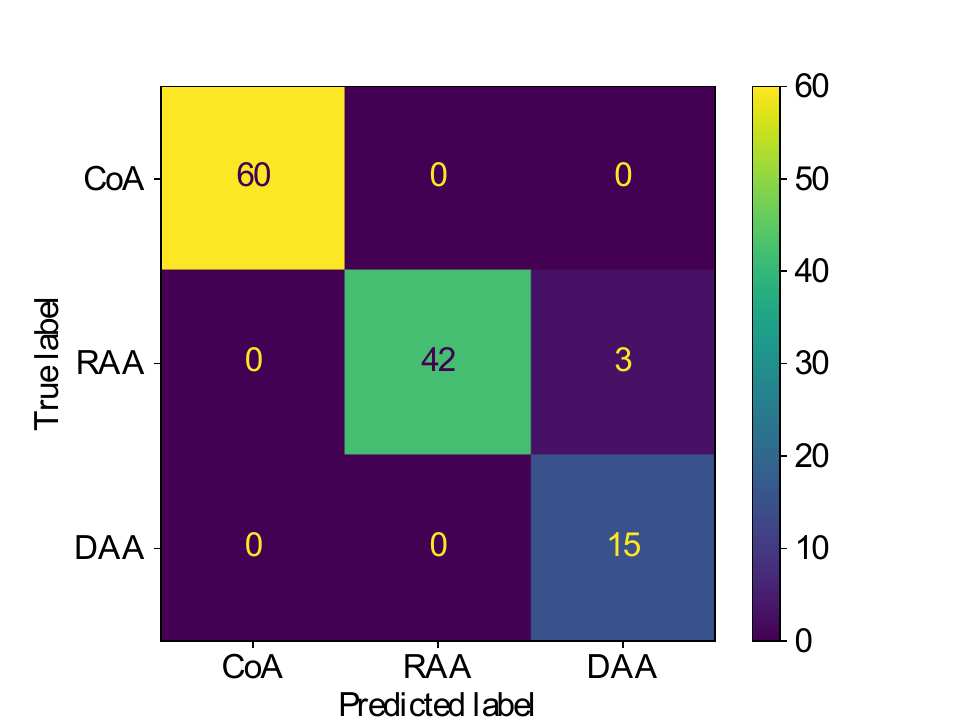}
         \caption{\textbf{DenseMulti} confusion matrix, three training rounds (1 misclassification per round).}
         \label{fig:densesegcm}
     \end{subfigure}
     \begin{subfigure}{0.49\textwidth}
         \centering
         \includegraphics[width=1.1\textwidth]{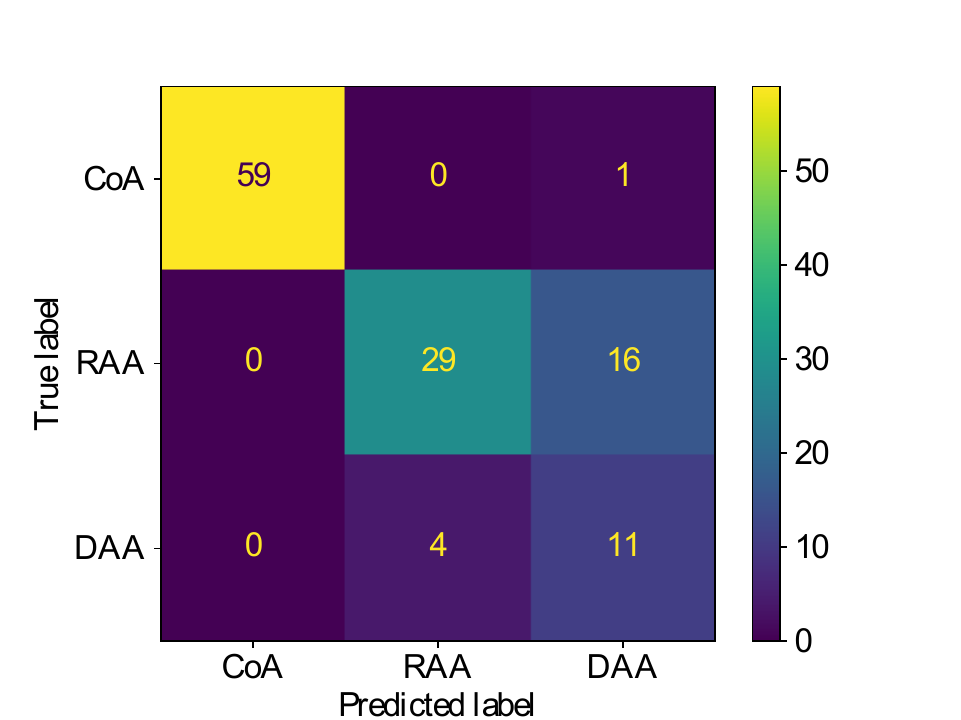}
         \caption{\textbf{DenseImage} confusion matrix, three training rounds.}
         \label{fig:densevolcm}
     \end{subfigure}
     
    \begin{subfigure}{0.49\textwidth}
         \centering
         \includegraphics[width=1.1\textwidth]{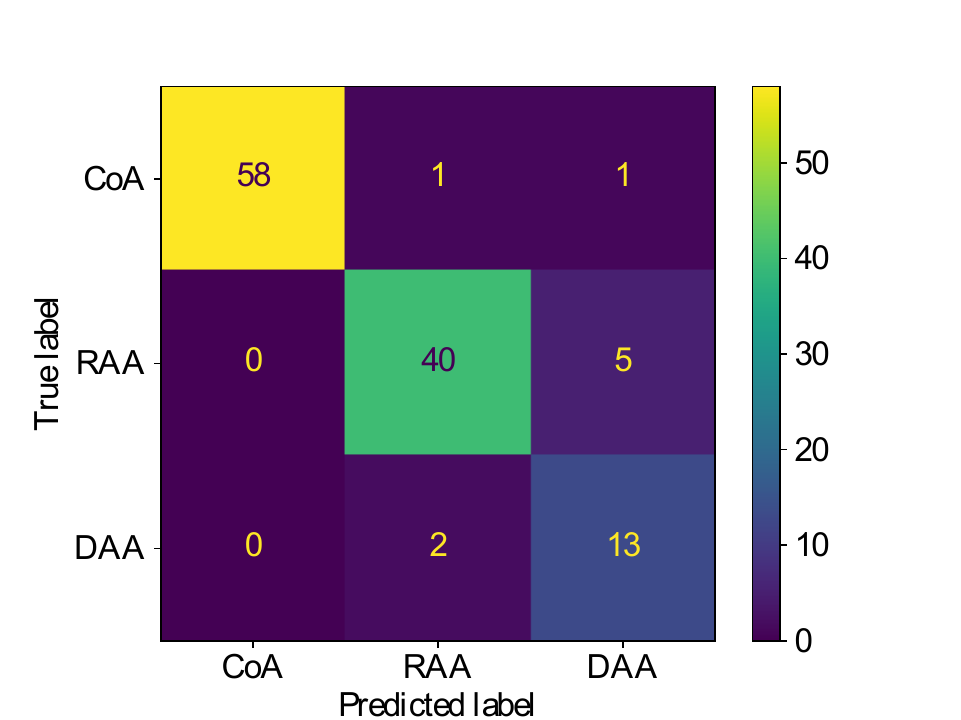}
         \caption{\textbf{DenseBin} confusion matrix, three training rounds.}
         \label{fig:densebin}
    \end{subfigure}
    \hfill
    \begin{subfigure}{0.49\textwidth}
         \centering
         \includegraphics[width=1.1\textwidth]{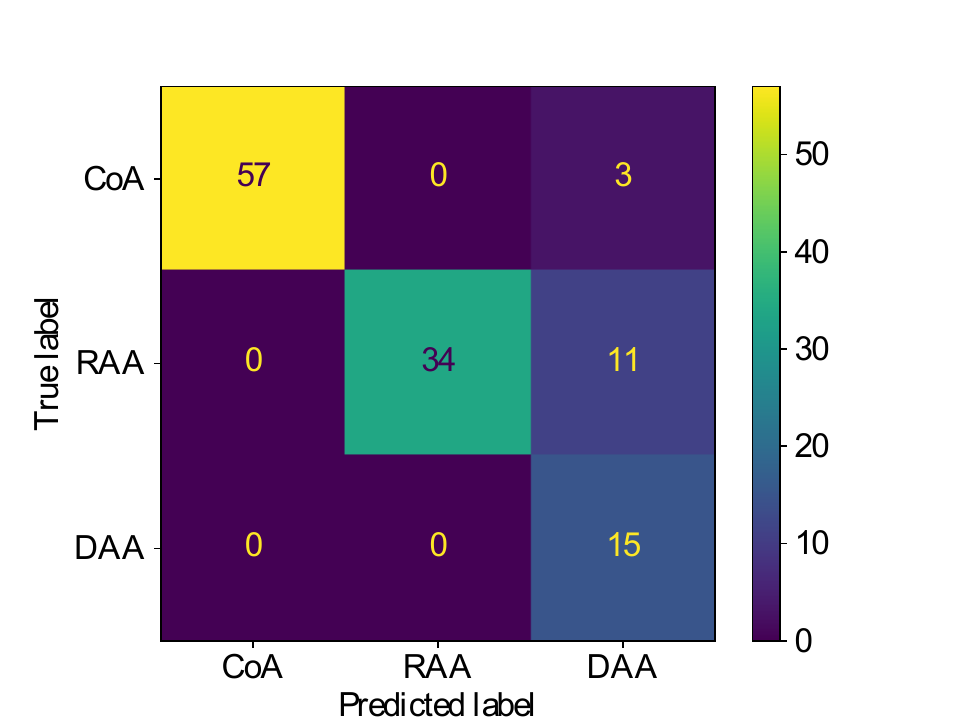}
         \caption{\textbf{DenseBin} before joint training confusion matrix, three training rounds.}
         \label{fig:densebin_before}
    \end{subfigure}
    
    \begin{subfigure}{0.49\textwidth}
         \centering
         \includegraphics[width=1.1\textwidth]{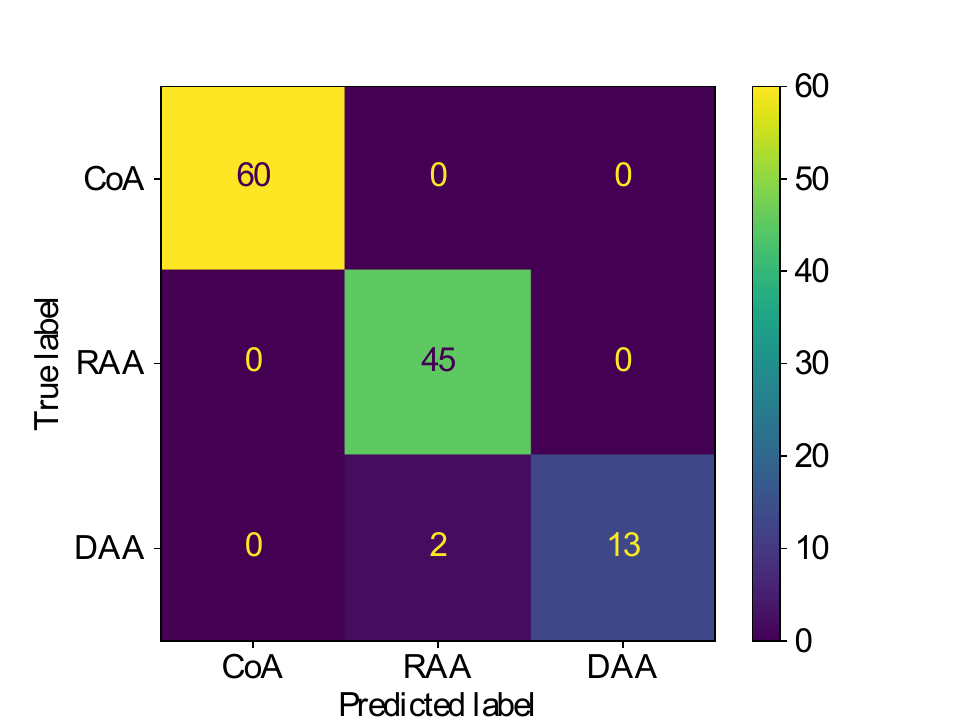}
         \caption{\textbf{DenseMulti} before joint training confusion matrix, three training rounds.}
         \label{fig:denseseg_before}
    \end{subfigure}
     
\end{figure}

Figs.~\ref{fig:densesegcm}, \ref{fig:densevolcm}, \ref{fig:densebin}, \ref{fig:denseseg_before} and \ref{fig:densebin_before} contain our classifier test set confusion matrices for our three training rounds.

\subsection{Qualitative analysis extended}
\label{appendix:qualitative_extended}

\begin{figure}[h]
     \centering
         \includegraphics[width=\textwidth]{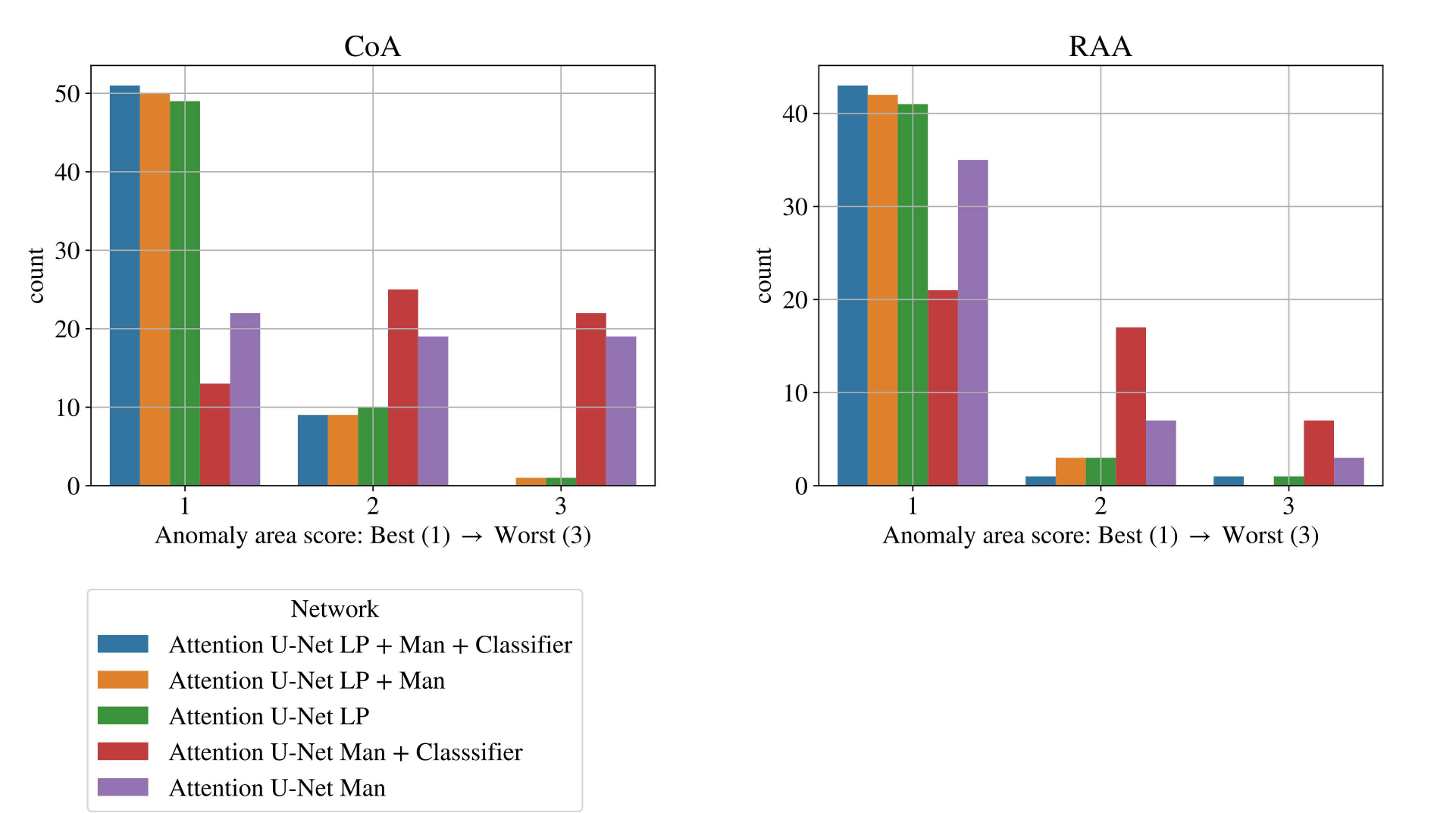}
         \caption{CoA  (left) and RAA (right) test set topology scores.}
         \label{fig:coa_topology}
\end{figure}

Here we extend our investigation into the lower-scoring subjects of our final framework (\textit{Attention U-Net LP + Man + Classifier} with \textbf{DenseMulti} classifier), as assessed by our qualitative analysis (Sec.~\ref{sec:results_qualitative}). We include our topology scores separated for CoA and RAA in Figs.~\ref{fig:coa_topology}.

\begin{figure}[h]
    \centering
    \includegraphics[width=\textwidth]{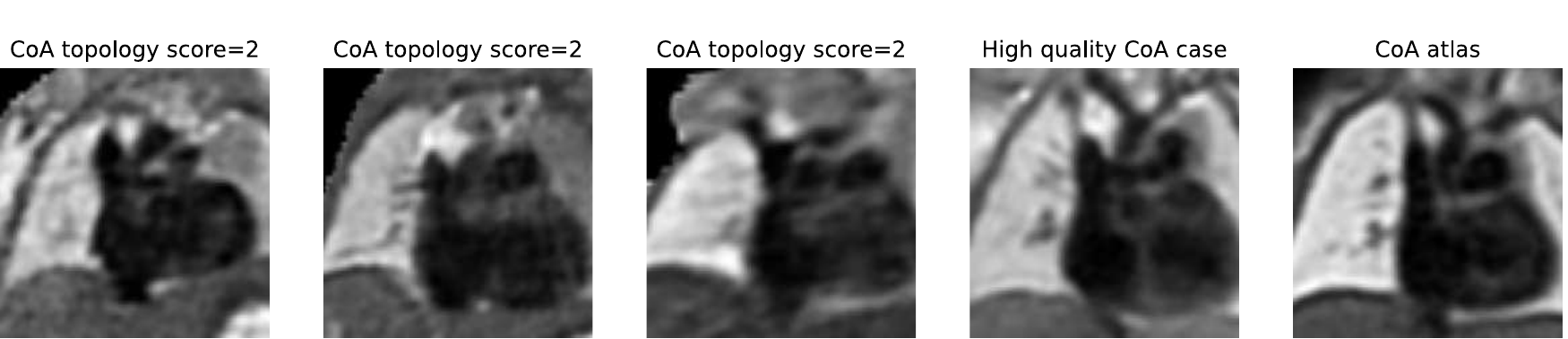}
    \caption{CoA test set cases with a topology score of 2 in our final framework, alongside a high-quality example case and our CoA atlas.}
    \label{fig:coa_scores2}
\end{figure}

In total, we find ten nine subjects with a topology score of 2 for the anomaly area, and one with a score of 3. These subjects comprise our two misclassified RAA cases (due to a partially and fully segmented double arch), as well as three CoA cases repeated across all three rounds. The latter cases include oversegmentation of the aorta into head and neck vessels, and merging into AD. We inspect the images and find lower quality and alignment compared to the atlas. Therefore we conclude that these subjects are outliers due to lower image quality, which is supported by the fact that three independently trained networks present segmentation issues on these subjects. Fig.~\ref{fig:coa_scores2} showcases the three CoA cases with a score of 2, alongside a high-quality correctly segmented case, and our CoA atlas to highlight the image quality differences.

\subsection{Latent space representations}
\label{appendix:latent}

\begin{figure}[h]
    \centering
    \includegraphics[width=\textwidth]{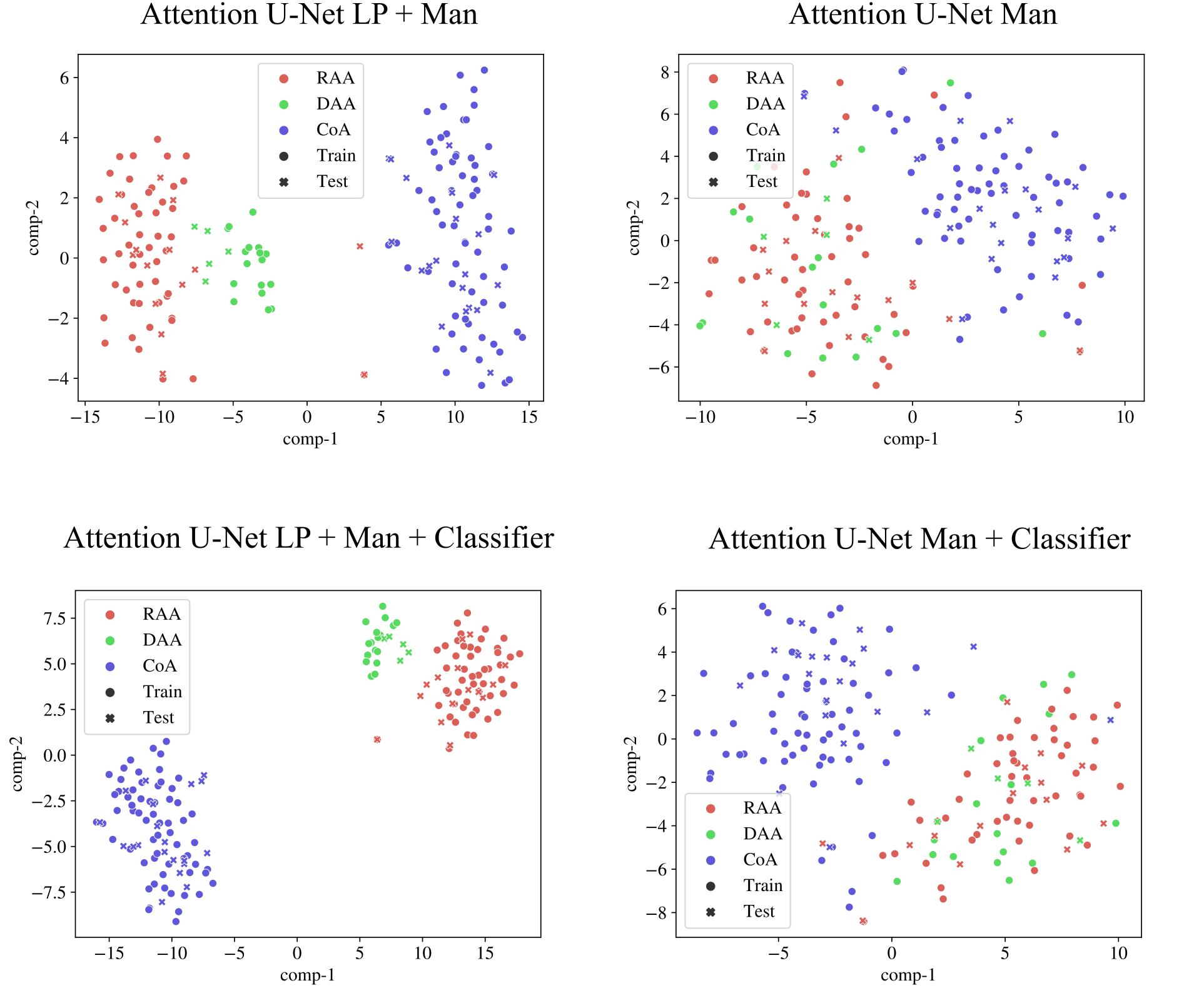}
    \caption{t-SNE visualisation of our network bottleneck features before (top row) and after (bottom row) the addition of an anomaly classifier. This is an example of one of our training rounds. }
    \label{fig:latent_classifier_3rounds}
\end{figure}

Fig.~\ref{fig:latent_classifier_3rounds} contains our t-SNE reduced latent space visualisations before and after adding a classifier to our segmentation frameworks, for one of our training rounds. We display both our binary segmentation networks (\textit{Attention U-Net Man}), and our proposed multi-class approach (\textit{Attention U-Net LP + Man}). We observe slightly further clustering after joint training with the anomaly classifier. However, this is not as pronounced as training using multi-class labels as opposed to binary manually segmented labels.

\end{document}